\documentclass[aps,twocolumn,superscriptaddress,prb,floatfix,citeautoscript,preprintnumbers]{revtex4-1}
\usepackage{graphicx,color}
\usepackage{amssymb}
\usepackage{amsmath}
\usepackage{epstopdf}
\usepackage[normalem]{ulem}
\usepackage{comment}
\usepackage{breakcites}
\usepackage[breaklinks=true]{hyperref}
\usepackage{subfigure,epsfig}

\usepackage{bbold}
\def\Id{\mathbb{1}}

\newcommand{\beq}{\begin{equation}}
\newcommand{\eeq}{\end{equation}}
\newcommand{\bal}{\begin{align}}
\newcommand{\eal}{\end{align}}

\newcommand{\nn}{{\nonumber}}

\newcommand{\ket}[1]{\mbox{$ | #1 \rangle $}}
\newcommand{\bra}[1]{\mbox{$ \langle #1 | $}}

\begin{document}

\title{Dynamics of quantum information in many-body localized systems}

\author{M. C.  Ba\~nuls}
\affiliation{Max-Planck-Institut f\"{u}r Quantenoptik, Hans-Kopfermann-Str. 1, D-85748 Garching, Germany}
\author{N. Y. Yao}
\affiliation{Physics Department, University of California Berkeley, Berkeley, California 94720, USA}
\author{S. Choi}
\affiliation{Physics Department, Harvard University, Cambridge, Massachusetts 02138, USA}
\author{M. D. Lukin}
\affiliation{Physics Department, Harvard University, Cambridge, Massachusetts 02138, USA}
\author{J. I. Cirac}
\affiliation{Max-Planck-Institut f\"{u}r Quantenoptik, Hans-Kopfermann-Str. 1, D-85748 Garching, Germany}

\preprint{NSF-ITP-17-088}
\date{\today }

\begin{abstract}
We characterize the  information dynamics of strongly disordered systems using a combination of analytics, exact diagonalization, and
matrix product operator simulations.
More specifically, we study the spreading of quantum information in three different scenarios: thermalizing, Anderson localized, and many-body localized.
We qualitatively distinguish these cases by quantifying the amount of remnant information in a local region.
The nature of the dynamics is further explored by computing the propagation of mutual information with respect to varying partitions. 
Finally, we demonstrate that classical simulability, as captured by the magnitude of MPO truncation errors, exhibits enhanced fluctuations near the localization  transition, suggesting the possibility of its use as a  diagnostic of the critical point. 
\end{abstract}

\maketitle

\section{Introduction}
\label{sec:intro}
In thermalizing quantum systems, it is typically thought that the microscopic information associated with any initial state is  lost as the system relaxes toward  equilibrium.
Even in the case of isolated systems (i.e.~absent a bath), information about an initial state is quickly spread over the entire system, rendering \emph{local} measurements incapable of any meaningful reconstruction \cite{Deutsch_1991_ETH,Srednicki_1994_ETH}.
However, strong disorder giving rise to localization can prevent equilibration, leading to ``memory'' of the initial state even at late times. 
Typically, this memory is associated with a lack of transport, implying for example, that microscopic information about the positions of particles remains at infinitely long times. 
While originally introduced by Anderson for the case of non-interacting systems~\cite{Anderson:2011wp}, more recently, it has been demonstrated that localization can  persist even in strongly interacting systems, leading to  a new dynamical phase of matter dubbed many-body localization (MBL)~\cite{PhysRevB.21.2366_1980,
basko2006,
PhysRevB.75.155111_2007,
znidaric08xxz,
pal2010,
bardarson12unbounded,
serbyn13slowgrowth,
serbyn13local,
Serbyn2014echoMBL,
Serbyn_2014_quench,
choi2015qcmbl,
Yao2015MBL_state_transfer,
Vosk_2014,
PhysRevLett.113.107204_2014,
PhysRevX.4.011052_2014,
huse14phenomenology,
luitz2015,
quantum_revivals_Vasseur_2015,
nandkishore15review,
vosk15prx,
agarwal15grif,
Singh_numerics_2015,
schreiber15obs,
imbrie16proof,
Vasseur_2016,
kondov2015exp,
choi2016two,
smith2016mbl}.

From an information theoretical perspective, the ability of localized systems to skirt thermalization suggests their use as a possible memory resource. In the many-body case, this paints a particularly intriguing picture where locally addressable degrees of freedom may emerge from a strongly-interacting system \cite{serbyn13local,Serbyn2014echoMBL,Serbyn_2014_quench,choi2015qcmbl,Yao2015MBL_state_transfer,huse14phenomenology}. However, the presence of strong interactions also leads to a logarithmic growth of entanglement entropy, manifesting as an intrinsic, slow dephasing mechanism~\cite{bardarson12unbounded,serbyn13slowgrowth,serbyn13local,Serbyn2014echoMBL,Serbyn_2014_quench,huse14phenomenology}; crucially, this implies that  not \emph{all} microscopic information of an initial state survives.

In this article, we perform a numerical study of the information dynamics in localized systems. 
By encoding a single qubit
of information in a local region, we quantify the amount of remnant information as a function of time, using the distinguishability of many-body density operators. 
Moreover,  we characterize the infinite temperature dynamics of 
the mutual information between a region and its complement (which extends the notion of entanglement entropy from zero to finite temperature, and satisfies an area law for any local Hamiltonian  in thermal equilibrium~\cite{wolf2008area})
using matrix product operator (MPO) simulations  \cite{verstraete04mpdo,zwolak04mpo,pirvu10mpo,Pollmann_2016_MPOMBL}.
Finally, we  find that ``classical simulability'' may serve as a diagnostic of the localization phase transition. 

Our paper is organized as follows.
In section~\ref{sec:model}, we begin by introducing the random field XXZ model and considering  two simple limits: the free, non-interacting case and the many-body localized $l$-bit Hamiltonian \cite{}. 
In addition, we outline the numerical tools used in the remainder of the paper.
In section~\ref{sec:memory}, we investigate the spreading of locally encoded information in both Anderson localized and MBL systems.
For Anderson localized systems, a  \emph{quantum} bit can be stored and trivially recovered; in this case, each  localized degree of freedom forms an independent, isolated qubit.
For MBL systems, these local qubits slowly entangle and dephase with one another, suggesting naively 
that only classical information survives at late times.
But this naive picture ignores the underlying mechanism of MBL dephasing. In particular, performing a simple local spin echo protocol leads to a revival of the quantum information and enables the recovery of a single qubit even at infinite temperatures for asymptotically long times \cite{huse14phenomenology,Serbyn2014echoMBL,quantum_revivals_Vasseur_2015}. 
Here, we consider a complementary scenario where one would encode multiple qubits,  each in a region of size $\ell$. In this case, the straightforward application of local spin-echo protocols does not lead to information recovery since dephasing still occurs between the qubits themselves \footnote{One can consider generalized spin echo protocols using orthogonal array techniques \cite{}, but a simple application of such a strategy leads to an exponential number of pulses (owing to multi-body interactions in the phenomenological MBL Hamiltonian \cite{}). }. 
To this end, we quantify the amount of information remaining in the region $\ell$; this amounts effectively to 
first tracing out the remainder of the system, and then asking for the maximum local recovery fidelity.
In section~\ref{sec:mutualinfo}, we highlight the use of infinite temperature MPO simulations to explore the dynamics of mutual information. 
By investigating the truncation errors associated with this tensor network simulation as a function of disorder strength, we find that  the classical simulability may serve as a diagnostic of the localization phase transition. This is elaborated upon in detail in  section~\ref{sec:errors}.
Finally, we conclude by summarizing our results in section~\ref{sec:conclu}.

\section{Model, Setup and Integrable Limits}
\label{sec:model}

We consider a one dimensional quantum spin chain of finite length $N$ with Hamiltonian: 
\beq
H=\sum_{i=0}^{N-2} J \left(S_i^x  S_{i+1}^x +S_i^y  S_{i+1}^y +\alpha S_i^z S_{i+1}^z\right) +\sum_{i=0}^{N-1}h_i S_i^z,
\label{eq:H}
\eeq
where $S_i^\mu$ ($\mu\in\{x,y,z\}$) are Pauli spin-1/2 operators acting on particle $i$, $J=1$ is the interaction strength between nearest neighbors, $h_i$ is a random on-site disorder field drawn from a uniform distribution $h_i\in[-h,h]$, and $\alpha$ is a dimensionless parameter  characterizing the XXZ anisotropy.
When $\alpha=0$, the model reduces to the non-interacting XY chain with random transverse field, 
which exhibits single particle localization for any value $h>0$ in the limit $N\rightarrow \infty$.
When $\alpha =1$, the spins couple via Heisenberg interactions, and the system undergoes an MBL phase transition with $h_c \approx 3.5$~\cite{pal2010,luitz2015,Singh_numerics_2015}.

As previously described, we are interested in situations where spin echo protocols are not applicable; to this end, we focus on the dynamics that lead to the spreading of information for an initial state, with one bit of information encoded in the leftmost spin. The remainder of the chain is prepared in the maximally mixed state (e.g.~infinite temperature):
\beq
\rho_{\varphi} = |\varphi\rangle\langle \varphi| \otimes  \left(\Id/2\right)^{\otimes {N-1}},
\label{eq:rhos}
\eeq
where $|\varphi\rangle$ describes the encoded pure state.
We will consider two types of states:
 $|\varphi\rangle=|X\pm\rangle$ or $\ |Z\pm\rangle$, which correspond to the eigenstates of $S_0^x$ and $S_0^z$.
 As we will see later, they can be associated to quantum and classical information. 

\subsection{Non-interacting and $l$-bit limit}
\label{subsec:xy}

Using the Jordan-Wigner transformation~\cite{jordanwigner1928}, the model can be written as a quadratic, free-fermion Hamiltonian, $H=\sum c^{\dagger}_p M_{pq} c_q$
where $M_{pq}\equiv-2 J (\delta_{p,q+1}+\delta_{p,q-1})+2 h_p \delta_{pq}$ and $c_i^\dagger$ ($c_i$) are creation (annihilation) operators at position $i$.
The eigenstates of the system can be simply described by  Slater determinant states of non-interacting single particle eigenmodes $b_k$, which can be  computed by diagonalizing the matrix $M=U\Lambda U^{\dagger}$, with $b_k=\sum_{l} U_{lk}^*c_l$. Likewise, the spectrum of the system is completely determined from the single particle energy eigenvalues $\Lambda_k$.
For any non-vanishing strength of the random magnetic field, each single particle eigenstate $k$ is spatially localized near some position $k_0$ with a characteristic localization length $\xi$ and $|U_{lk}|^{2}\propto e^{-|l-k_0|/\xi}$.

Starting from Eq.~\eqref{eq:rhos} with  $|\varphi\rangle=|X\pm\rangle,\ |Z\pm\rangle$, the time evolved
density matrix is
\begin{align}
\rho_{Z_{\pm}}(t)=&\frac{1}{2^{N-1}}
\sum_{r=0}^{N-1} 
\left \{ |V_r|^2 \frac{1\pm\sigma_r^z}{2}  \right.  \label{eq:rhoZ_spin} \\
&\mp
\left.\sum_{s>r} \left ( V_r V_s^* \sigma_r^- \otimes \sigma_{r+1}^z \otimes\ldots \otimes \sigma_{s-1}^z \otimes \sigma_s^+ \right. \right. \nn \\
&
\left.
\phantom{\sum_{s>r} } \left. 
+
V_r^* V_s \sigma_r^+ \otimes \sigma_{r+1}^z \otimes\ldots \otimes \sigma_{s-1}^z \otimes \sigma_s^-
\right)
\right \}, \nn
\end{align}
\begin{align}
\rho_{X\pm}(t)=\frac{1}{2^N}\left \{ \Id \pm \sum_{r=0}^{N-1}\left( V_r \sigma_0^z \otimes\ldots \otimes \sigma_{r-1}^z \otimes \sigma_r^- +h.c. \right)\right \},
\label{eq:rhoX_spin} 
\end{align}
where $V_r (t) = \sum_{l=0}^{N-1}U_{0l}e^{i\Lambda_l t}U_{lk}^{\dagger}$ is the quantum amplitude of a single particle propagating from the $0$-th to $r$-th site over time $t$.

For $\alpha \neq0$, the model is interacting and there is no generic analytic solution. However, when the interacting system is in the MBL phase, one can use a phenomenological Hamiltonian of the form~\cite{serbyn13local,huse14phenomenology},
\begin{align}
H_{\mathrm{eff}}=&\sum_{i} \epsilon_i \tau_z^{[i]}+\sum_{i,j} K_{ij}^{(2)} \tau_z^{[i]} \tau_z^{[j]} 
 +\dots,
\label{eq:lbits}
\end{align}
where $\tau_z^{[i]}$ are Pauli operators corresponding to so-called $\l$-bits (logical bits), and $K_{i_1,i_2,...}^{(M)}$ are the coefficients of $M$-body interactions that decay exponentially in space 
and with the number of participating  $\l$-bits, $M$.
The $\tau^{[i]}$ operators are related by a quasi-local unitary to the original spin degrees of freedom and thus, are localized around a physical site $i$~\cite{chandran16dims,ros15integrals}.
In the remaining sections, we will utilize this description to estimate the qualitative behavior of information dynamics in the  MBL phase.

\subsection{Numerical method based on tensor networks}
To probe system sizes larger than those possible via exact diagonalization, we will consider approximate numerical simulations based upon tensor network (TNS) techniques  using MPO~\cite{verstraete04mpdo,zwolak04mpo,pirvu10mpo}.
An MPO  is a particular tensor network ansatz for operators. For a spin chain of $N$ sites,  MPO's take the form
\begin{align}
   O = 
         \sum_{\{i_k,j_k\}} 
         {\rm Tr} \left[ M[0]^{i_0j_0} \right. & \left. \cdots M[N-1]^{i_{N-1}j_{N-1}} \right]  \nonumber \\
         & \times |i_0\ldots i_{N-1}\rangle\langle j_0\ldots j_{N-1}|,
\end{align}
where $i_k,\, j_k\in\{0,\, 1\}$ label the physical basis of the $k$-th site, and each $M[k]^{ij}$ is a $D\times D$ matrix.
Such an MPO can be used as an ansatz to represent the state of a quantum many-body system.
Using standard tensor network techniques \cite{verstraete08algo}, it is thus possible to simulate the time evolution of
a mixed state, which in the case of Eq.~\eqref{eq:rhos} has  an exact MPO representation with bond dimension, $D=1$. 
In particular, we can find an MPO approximation to the time evolved state of the system, $\rho(t)$,
by successively applying small steps of evolution onto the initial state, and truncating the result to a maximum allowed 
bond dimension.

We note that, in the non-interacting case, $\rho_{Z_{\pm}}(t)$ and $\rho_{X_{\pm}}(t)$ in Eq.~\eqref{eq:rhoZ_spin}~and~\eqref{eq:rhoX_spin} can be exactly written as MPOs of bond dimensions $D=4$ and $D=2$, respectively 
(see appendix~\ref{app:XY}),
already indicating that this non-interacting case can be efficiently simulated using this method.
In a generic interacting case, a larger bond dimension is required as the system evolves in time, and accurate simulations become more difficult.
However, the computational cost scales polynomially with the system size, enabling one to explore 
considerably longer chains than with exact diagonalization.
On the other hand, the cost grows linearly with the number of evolution steps applied, so that the total  time 
that can be simulated with this method is still limited.
More importantly, the truncation of the bond dimension introduces a numerical error, which may grow fast with time 
if the bond dimension required for a precise description of the true evolved state increases rapidly. 
As discussed below, the minimal value of $D$ necessary for a certain precision depends on the Hamiltonian 
parameters and the required $D$ is typically small if the system is in the localized phase~\cite{znidaric08xxz}. 

\section{Information spreading and dynamics}
\label{sec:memory}
As previously discussed, locally encoded information in MBL systems may dephase over time owing to weak interactions, and if any part of the system is traced out, the information can be lost.
To make this statement more precise,
we focus on a scenario in which 
the leftmost spin starts out in a pure state, $\ket{\varphi}$, the system evolves for time $t$, 
and afterwards the rightmost $L-\ell$ spins are traced out.
We compute the distinguishability for pairs of 
states that were initially oppositely polarized  along  
either $\hat{x}$  or $\hat{z}$ direction. 
This quantity measures the optimal probability of 
distinguishing
whether we started with $+$ or $-$ polarization. 
It is computed as 
$\frac{1}{2}\| {\cal{D}}_{t,\ell}(\sigma_{\alpha})\|_1=\frac{1}{2}\| {\cal{D}}_{t,\ell}(\ket{\alpha +}\bra{\alpha +})-{\cal{D}}_{t,\ell}(\ket{\alpha -}\bra{\alpha -})\|_1$ ($\alpha=x,\, z$),
where 
${\cal{D}}_{t,\ell}(|{\varphi}\rangle\langle\varphi|)=\mathrm{tr}_{L-\ell}\left [ U(t)|\varphi\rangle\langle \varphi| U(t)^{\dagger} \right ] $
describes the unitary time evolution of the initial state followed by tracing over the rightmost $L-\ell$ spins.
Notice that  $0\leq \| {\cal{D}}_{t,\ell}(\sigma_{\alpha})\|_1\leq 2$,
with the maximum value 2, when the states are perfectly distinguishable (e.g. at $t=0$), and
the minimum value 0, when both states are indistinguishable and the information is completely lost.
For very large magnetic field, we might expect a large value of $\| {\cal{D}}_{t,\ell}(\sigma_{z})\|_1$,
even in the interacting case, and we may relate distinguishability for the $\hat{z}$ direction to the classical information  that can be stored. 
On the other hand,  $\| {\cal{D}}_{t,\ell}(\sigma_{x})\|_1$, which we relate to the quantum information, may be very different.
As we will show, it may happen that the $x$-distinguishability vanishes, while the $z$-distinguishability remains close to maximal, meaning that we can retrieve a classical but not a quantum bit.

Furthermore, the distinguishabilities along three orthogonal directions also allow us to compute an upper bound to the 
recovery fidelity of our protocol.
To this end, after the evolution, a trace preserving completely-positive map (ie., a physical action), $\cal{R}_{\ell}$, can be 
applied to the $\ell$ leftmost spins in order to recover the encoded information.
The fidelity of the procedure is defined as the average fidelity between initial and final states,
\beq
F({\cal{R}}_{\ell},t,\ell)= \int d \mu_{\varphi} \langle \varphi| {\cal{R}}_{\ell}\left [{\cal{D}}_{t,\ell} (|\varphi\rangle\langle\varphi|) \right]|\varphi\rangle, 
\label{eq:RF}
\eeq
averaging over all initial pure states of the qubit with the standard measure.
The optimal recovery fidelity, $F(t,\ell)=\sup_{\cal{R}_{\ell}} F({\cal{R}}_{\ell},t,\ell)$, corresponds to an optimization over the quantum operation, $\cal{R}_{\ell}$, which 
in general is difficult to solve.
Nevertheless, a strict upper bound can be computed from the distinguishabilities of pairs of 
completely polarized initial states of the qubit along three orthogonal 
directions~\cite{mazza12majorana},
\begin{align}
F(t,\ell)&\leq \frac{1}{2} +\frac{1}{12} \sum_{\alpha=x,y,z} \| {\cal{D}}_{t,\ell}(\sigma_{\alpha})\|_1.
\end{align}
 The symmetry of our problem ensures that $\| {\cal{D}}_{t,\ell}(\sigma_{y})\|_1=\| {\cal{D}}_{t,\ell}(\sigma_{x})\|_1$, so that 
 it is enough to consider the previously mentioned $\| {\cal{D}}_{t,\ell}(\sigma_{x})\|_1$ and $\| {\cal{D}}_{t,\ell}(\sigma_{z})\|_1$.

\begin{figure*}[t]
\includegraphics[width=0.24\textwidth]{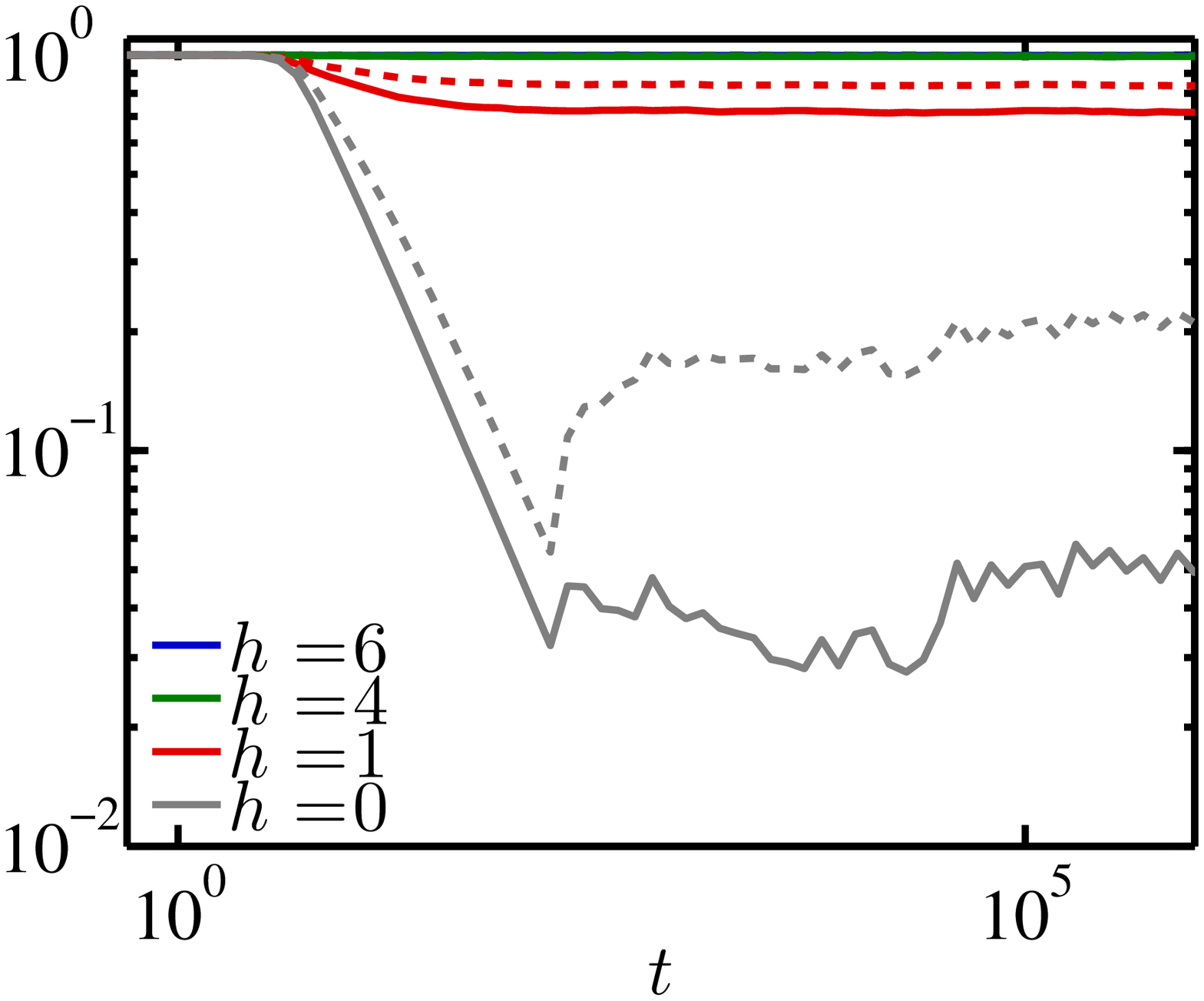}
\includegraphics[width=0.24\textwidth]{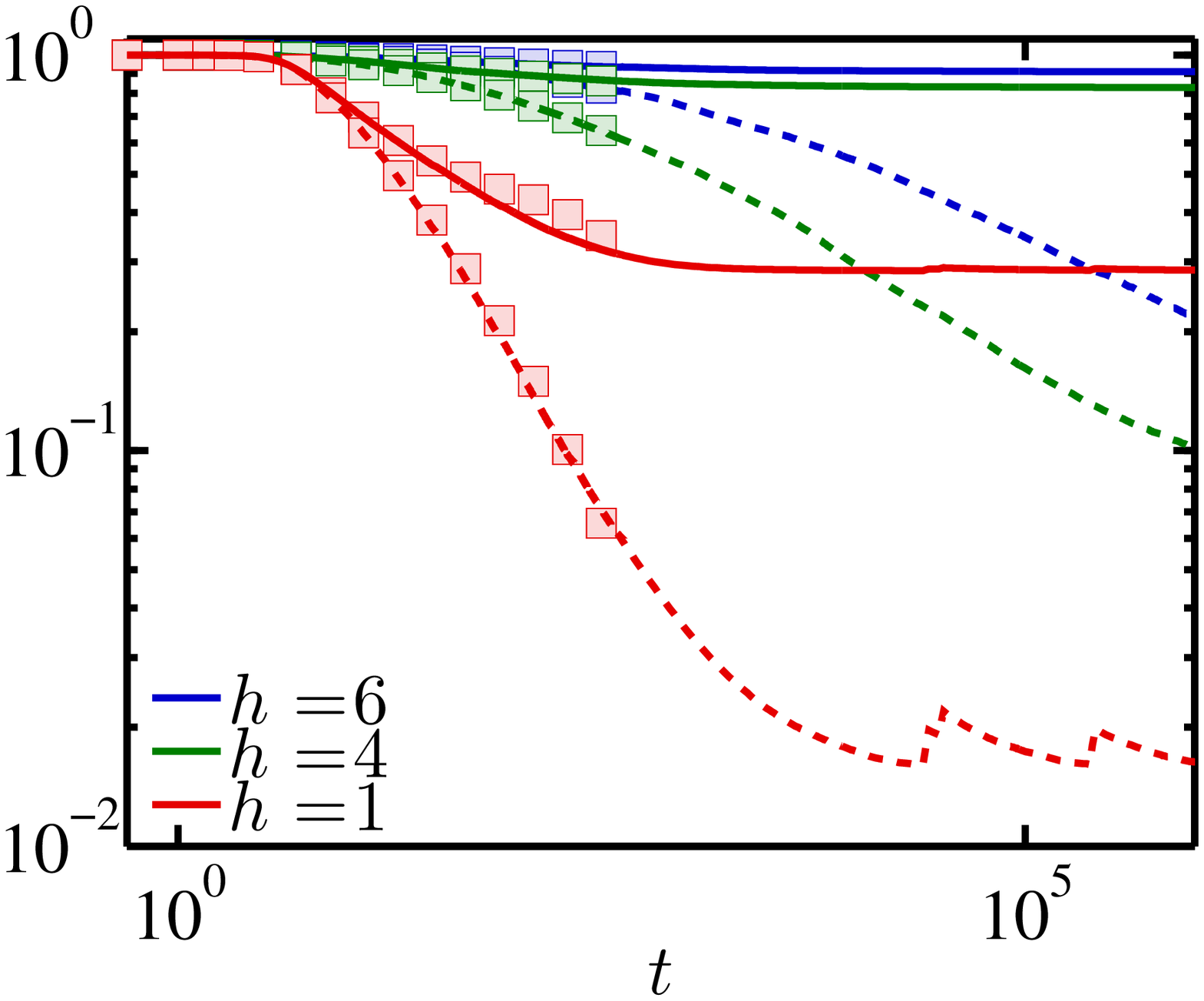}
\includegraphics[width=0.25\textwidth]{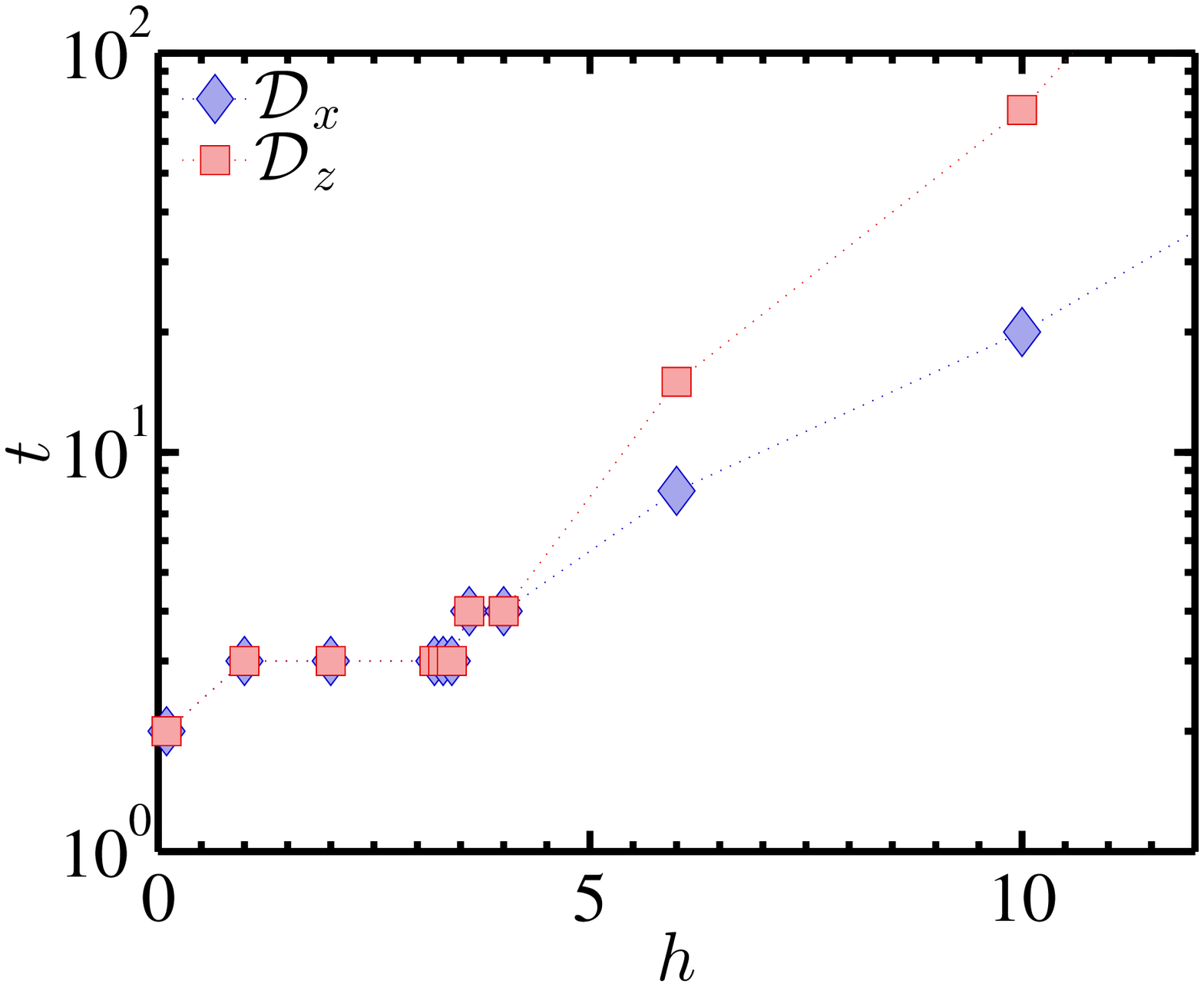} 
\caption{
Numerical results on the evaporation of information in disordered system.
\textbf{a-b} Disorder averaged distinguishabilities $\frac{1}{2}\|{\cal{D}}_{t,\ell=4}(\sigma_{z})\|_1$ (solid) and $\frac{1}{2}\|{\cal{D}}_{t,\ell=4}(\sigma_{x})\|_1$ (dashed lines). \textbf{a} Non-interacting case ($\alpha = 0$) for a chain of $L = 100$ sites.
The disorder strengths are $h=0$, $1$, $4$, and $10$ from bottom to up.
For each line, data has been averaged over $100$ disorder realizations. We show the time range before finite size effects appear. 
\textbf{b}
 Heisenberg interacting case ($\alpha = 1$), computed by exact diagonalization for $L=12$ sites (lines) and by MPO time evolution for $L=40$ (discrete data points).
The disorder strengths are $h=1$, $4$, and $6$ from bottom to up.
For each line, data has been averaged over $10$ disorder realizations.
\textbf{c}
Distinguishability lifetime as a function of disorder strength. Blue diamonds and red squares indicate the timescales under which the average distinguishability in $X$ and $Z$, respectively, for the average over $20$ instances, drops below a threshold value $0.99$, for a cut at $\ell=4$. 
}
\label{fig:distingXY}
\end{figure*}

\emph{Non-interacting case.}---
Using the exact solution in \eqref{eq:rhoZ_spin} and \eqref{eq:rhoX_spin}, the distinguishabilities can then be expressed as
\begin{align}
\label{eqn:non_int_D}
\| {\cal{D}}_{t,\ell}(\sigma_{z}) \|_1=2 {\sum_{r=0}^{\ell-1} |V_r|^2}\nn \;, \;\;
\|{\cal{D}}_{t,\ell}(\sigma_{x})\|_1 =2\sqrt{\sum_{r=0}^{\ell-1} |V_r|^2}.
\end{align}
From the unitarity  $\sum_{r=0}^{\ell-1} |V_r|^2 =1-\sum_{r=\ell}^{N-1} |V_r|^2  < 1$, we find $\| {\cal{D}}_{t,\ell}(\sigma_{z}) \|_1\leq \| {\cal{D}}_{t,\ell}(\sigma_{x}) \|_1$, implying that the distinguishability of 
$\sigma_x$ polarization is always better than that of 
$\sigma_z$ at all times and partitions.

When the system is disordered ($h>0$) the eigenmodes are exponentially localized.
We can then bound the probability that the initial particle propagates beyond the
$\ell$-th site at time $t$,  $\sum_{r=\ell}^{N-1}|V_r(t)|^2\leq C N (N-\ell) e^{-\ell/\xi}$ for some constant $C$, and since
 $\sum_{r=0}^{N-1}|V_r|^2=1$, we conclude that 
$\| {\cal{D}}_{t,\ell}(\sigma_{z}) \|_1 \geq 2-2 C N (N-\ell) e^{-\ell/\xi}$ and 
$\|{\cal{D}}_{t,\ell}(\sigma_{x})\|_1 \geq 2\sqrt{1-2 C N (N-\ell) e^{-\ell/\xi}}$.
Thus, in the non-interacting case information stored in both $\sigma_x$ and $\sigma_z$ remains localized at arbitrarily long times.

Figure~\ref{fig:distingXY}(a) illustrates the behavior of the disorder averaged 
$\frac{1}{2}\| {\cal{D}}_{t,\ell}(\sigma_{x})\|_1$ and $\frac{1}{2}\| {\cal{D}}_{t,\ell}(\sigma_{z})\|_1$
for a chain of length $L=100$ as a function of time with the cut at $\ell=4$. 
In the absence of disorder $h=0$, both values in Eq.~\eqref{eqn:non_int_D} decay after a certain time as  $t^{-1/2}$, before finite size effects are apparent. 
For disordered systems both quantities saturate within finite time.
The saturation times as well as the final values of distinguishabilities depend on the disorder strength.

\begin{figure}
\begin{center}
\includegraphics[width=0.46\columnwidth]{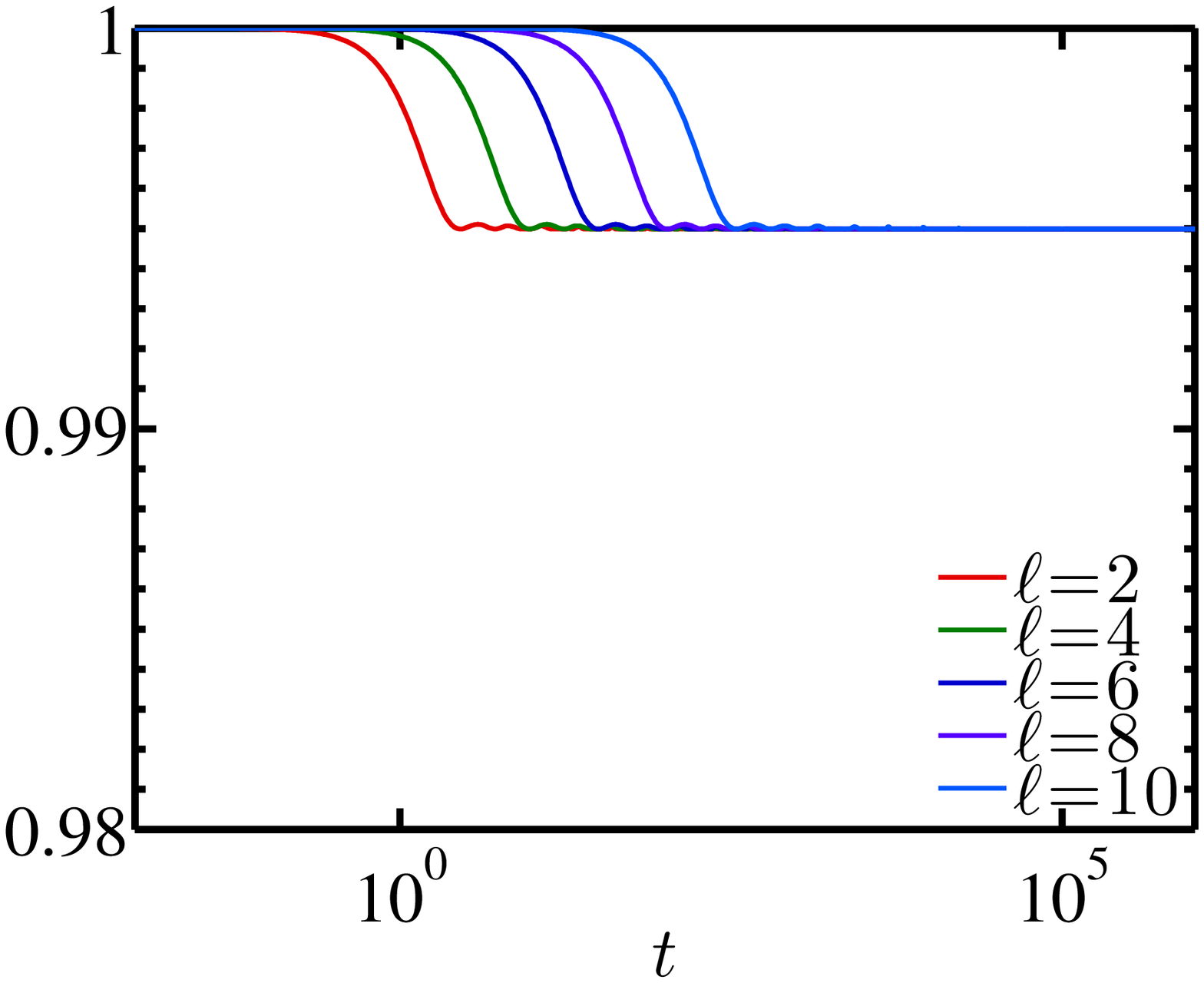} 
\includegraphics[width=0.45\columnwidth]{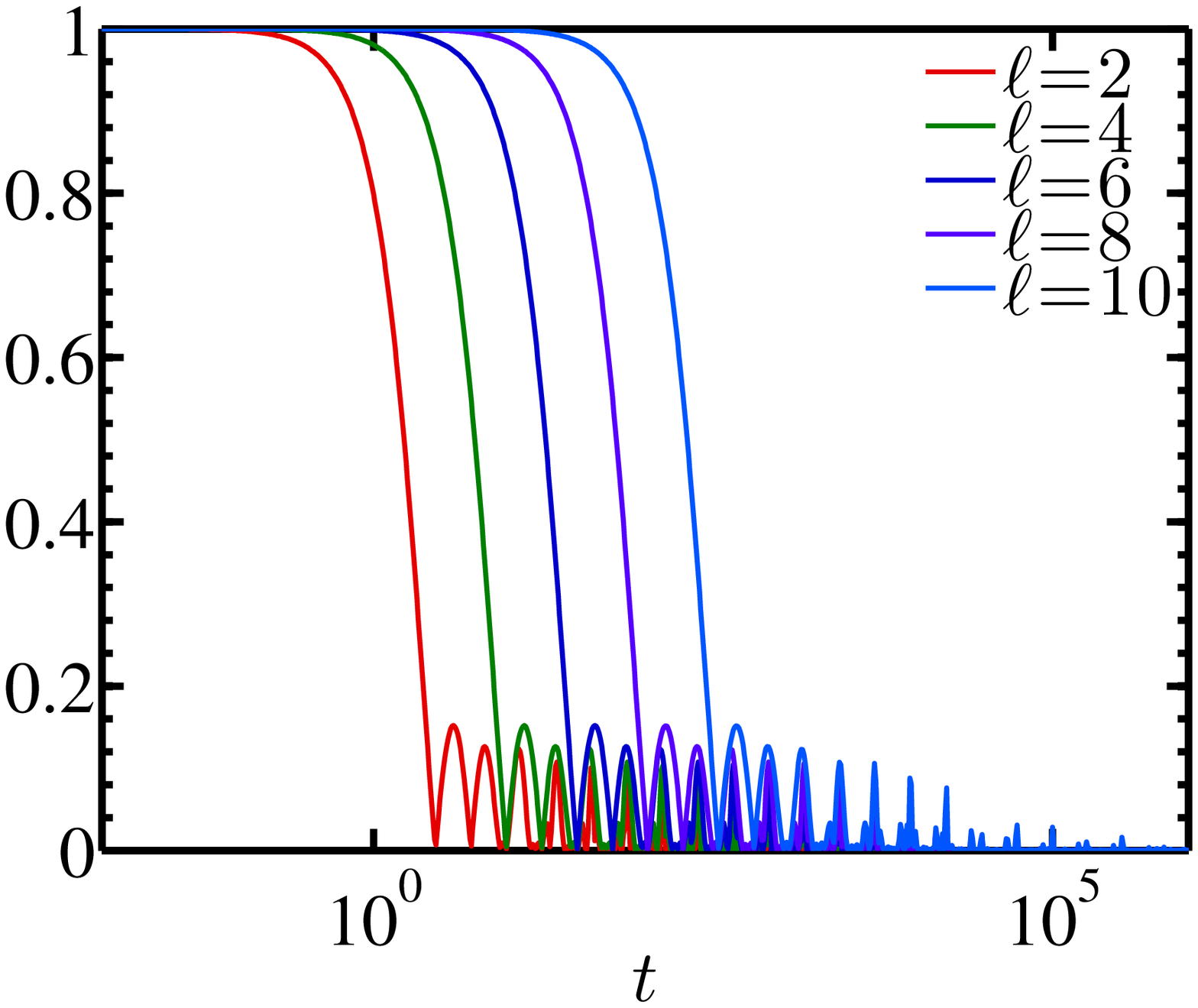}
\caption{Distinguishability of $Z$ (left) and $X$ (right) polarized states for different cuts, $\ell$, as a function of time as given by the phenomenological model~\eqref{eq:distingLbit} for a localization length $\xi=10$, $\beta=0.1$ and system size $N=100$.}
\label{fig:distingMBLtheory}
\end{center}
\end{figure}

\emph{Interacting case.}---
We first make use of the phenomenological $l$-bit model \eqref{eq:lbits}
to qualitatively predict the behavior of distinguishability in the MBL phase, assuming that the system is deep in the localized phase.
In such case, the physical spins and $l$-bits coincide, and the $\rho_{X\pm}$ states approximately correspond to a polarized first $l$-bit, $(\Id\pm\tau_x^{[0]})/2$,
and totally mixed states of the rest. 
Since the coefficients $K^{(m)}$ decay exponentially with the order, $m$, we consider, as a first order approximation, only terms with $m\leq2$.
Under these approximations, the time-dependent reduced density matrix for the 
first $\ell$ spins can be written as
 \begin{align}
 \tilde{\rho}_{\ell}(t)=\frac{1}{2^{\ell}}&\left(\Id \pm\prod_{k=\ell}^{N-1}\cos(2 t K_{0k}^{(2)})\right.\nn\\
 &\left.\left[\cos(2 t \epsilon_0) \tau_x^{[0]}+\sin(2 t \epsilon_0) \tau_y^{[0]}\right]\right).
 \end{align}
 For initial states $\rho_{Z\pm}$ we may assume the simplest first order decomposition 
 \beq
\sigma_z^{[0]}\approx \sqrt{1-\beta^2} \tau_z^{[0]}+\beta \tau_x^{[0]},
\label{eq:effecZ}
\eeq
where $\beta$ is a small dimensionless parameter characterizing the overlap between the physical spin operator $\sigma_z^{[0]}$ and the $l$-bit operator $\tau_x^{[0]}$~\footnote{Strictly speaking, owing to a symmetry of our Hamiltonian, total magnetization $\sum_i S_i^z$ is a conserved quantity. Hence, the leading order correction to $\sigma_z^{[0]}$ in $l$-bit basis need not be $\tau_x^{[0]}$. Here, we focus on a generic situation independent of  symmetry considerations for the purpose of understanding qualitative behaviors.}.
Correspondingly, the time-evolved density matrix becomes
\begin{align}
 \tilde{\rho}_{\ell}^{Z\pm}(t)=\frac{1}{2^{\ell}}&\left(\Id\pm \sqrt{1-\beta^2} \tau_z^{[0]} \pm \beta \prod_{k=\ell}^{N-1}\cos(2 t K_{0k}^{(2)})\right.\nn\\
 &\left.\left[\cos(2 t \epsilon_0) \tau_x^{[0]}+\sin(2 t \epsilon_0) \tau_y^{[0]}\right]\right).
 \end{align}
Thus, for the distinguishabilities, we can approximate 
\begin{align}
\|{\cal{D}}_{t,\ell}(\sigma_{x})\|_1 &\approx 2 |x(\ell,t)|,\nn\\
\|{\cal{D}}_{t,\ell}(\sigma_{z})\|_1 &\approx 2 \sqrt{1-\beta^2 (1-x(\ell,t)^2)}.
\label{eq:distingLbit}
\end{align}
 where we introduced $x(\ell,t)=\prod_{k=\ell}^{N-1}\cos(2 t K_{0k}^{(2)})$.
This parameter $x(\ell, t)$ characterizes the degree of dephasing of 
the first $l$-bit
induced by interactions with $l$-bits located further than $\ell$ sites. Initially, $x(\ell,t=0)=1$, but even in the deeply localized regime, this parameter may become smaller and explore all of its allowed value range, $0\leq x(\ell,t)\leq1$ at late times.
 For sufficiently large $\ell$, the  decrease of $x(\ell,t)$ occurs only at exponentially long time
 as
 $x(\ell,t)\approx 1-2 t^2 (N-\ell)e^{-2\ell/\xi}.$
From Eq.~\eqref{eq:distingLbit}, we conclude that in the MBL phase, the distinguishability of states that encode $\sigma_z$ remains lower bounded by  $2\sqrt{1-\beta^2}$.
In contrast,   $\|{\cal{D}}_{t,\ell}(\sigma_{x})\|_1$
will eventually vanish; all the information on the coherence will be lost (Fig.~\ref{fig:distingMBLtheory}).

We can compare these simple estimates with numerical results obtained from exact diagonalization for small system sizes and from MPO approximations for large system sizes.
Figure \ref{fig:distingXY}(b) illustrates the results obtained with ED for small chains of $L=12$, and with MPO for $L=40$ until time $t=400$.
As expected, simulation results show that in the interacting case, $\| {\cal{D}}_{t,\ell}(\sigma_{z})\|_1$ and $\|{\cal{D}}_{t,\ell}(\sigma_{x})\|_1$ exhibit qualitatively distinct behavior from the free fermionic case.
We observe that $\| {\cal{D}}_{t,\ell}(\sigma_{z})\|_1 \geq \| {\cal{D}}_{t,\ell}(\sigma_{x})\|_1$ at all times, and, moreover, the distinguishability of  $\sigma_x$ polarized states decays for all values of the disorder strength, even deep within the localized 
regime ($h=6$), at sufficiently long times.
The $\sigma_z$ distinguishability, on the other hand, seems to reach a plateau at higher value for increasing disorder strength, $h$. 
Although we observe similar plateaus even for relatively small values of the disorder, c.f. $h=1$, from more detailed comparison of different system sizes,
we attribute it to the effects of finite system sizes, most important for the weakest disorder. 
Such effects are evident even at moderate times $t=\mathcal{O}(100)$, where also the MPO approximate results for $L=40$ deviate from the small chain results.

Based on this analysis, we conclude that in Anderson localized systems, information encoded in both $\sigma_x$ and $\sigma_x$ polarization may remain localized for an arbitrary long time, and can be, in principle, reconstructed even after some part of the system is traced out.
In contrast, however, in MBL systems, information encoded in $\sigma_z$ and $\sigma_x$ exhibit qualitatively distinct behaviors; while \emph{classical} information encoded in the $\sigma_z$ polarization (i.e. qubit states $|0\rangle$ or $|1\rangle$), remains localized, coherent \emph{quantum} information encoded in the $\sigma_x$ polarization (e.g. $|X\pm\rangle$ states) is inevitably delocalized and eventually lost by tracing out. 
Nevertheless, as shown in Eq.~\eqref{eq:distingLbit} for strong disorder, the dephasing of the coherence information requires an exponentially long time, suggesting that the spreading of the information still remains slow as predicted from previous works~\cite{bardarson12unbounded,serbyn13slowgrowth,serbyn13local,choi2015qcmbl}.
In Fig.~\ref{fig:distingXY}(c) we explicitly show the numerically extracted times for which $\sigma_x$ and $\sigma_z$ distinguishabilities are maintained above a certain threshold as a function
of the disorder.
We find that the spreading of the information is dramatically slowed down in increasing disorder strength.

\begin{figure}
\begin{center}
\includegraphics[width=0.45\columnwidth]{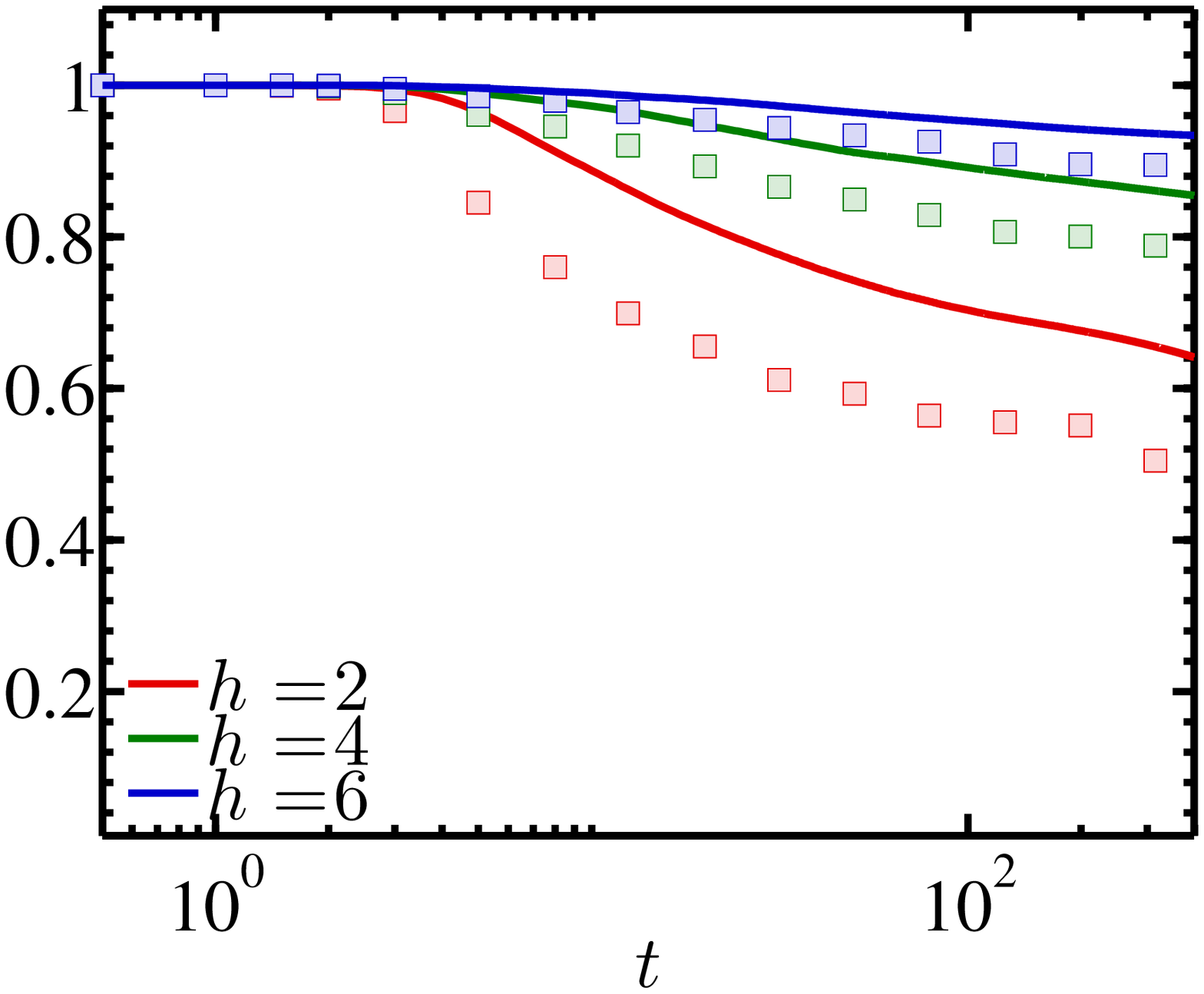} 
\includegraphics[width=0.45\columnwidth]{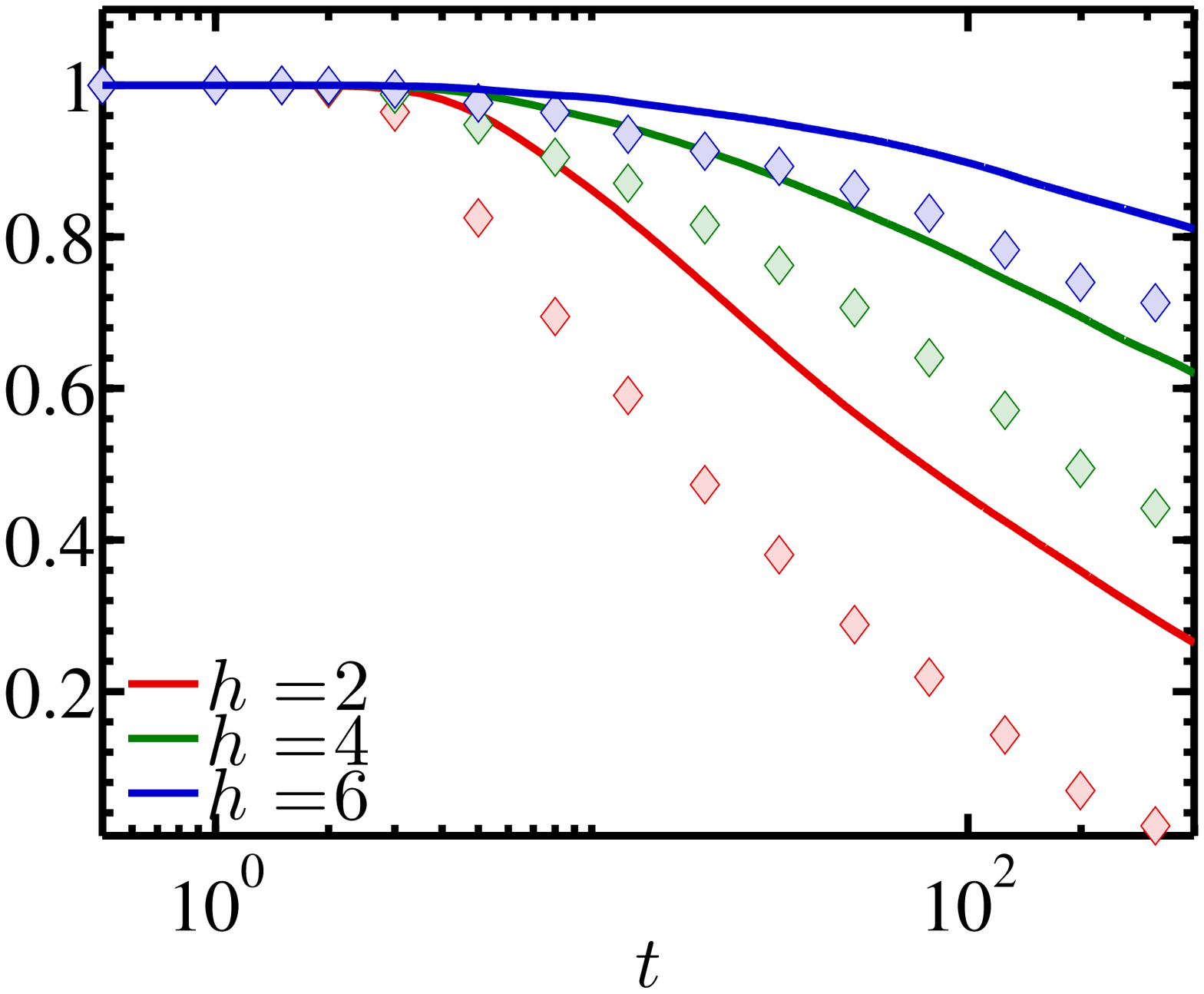}
\caption{Recovery of classical (left) and quantum (right) information, $\Delta m_{Z\pm}(t)$ (resp. $\Delta m_{X\pm}(t)$), for disorder strengths $h=2,\,4,\,6$, compared to the distinguishability of $z$ ($x$) polarized states (solid lines). All data correspond to MPO calculations on a chain of length $N=40$, using up to $D=80$ and averaging over $10$ instances..}
\label{fig:recoZdisting}
\end{center}
\end{figure}

While the previous analysis provides the bound on the amount of information that remains near the vicinity of the initial qubit, it does not necessarily mean that this information can be feasibly recovered from the local region via any practical protocol. In order to utilize the remnant information, one also needs to devise a recovery protocol that is independent from the encoded information.
Here, we consider one simple example based on time-reversed unitary evolution, ${\cal{R}_\ell} = U_{\ell}^\dagger(t) $, where $U_{\ell}(t)$ is the evolution operator for a time $t$ under the Hamiltonian restricted to the leftmost $\ell$ sites.
The measurement of the polarization on the first spin provides the information on encoded $\sigma_z$ polarization with the correlation
\begin{equation}
m_{Z\pm}(t)=\mathrm{tr}\left [\sigma_z^{[0]} U_{\ell}(t)^{\dagger}\mathrm{tr}_{L-\ell}\left (U(t) \rho_{Z\pm} U(t)^{\dagger}\right )U_{\ell}(t)\right ].
\label{eq:recovery}
\end{equation}
If $m_{Z\pm} = 2$, the correlation between encoded and decoded information is perfect, while
if $m_{Z\pm} = 0$, all the encoded information is lost in the decoded state.
By comparing the difference between $\Delta m_{Z}(t)=m_{Z+}(t)-m_{Z-}(t)$ and the distinguishability bound $\|\mathcal{D}_{t,\ell}(\sigma_z)\|_1$, one can estimate the performance of our protocol compared to the theoretical bound [see Fig. \ref{fig:recoZdisting}].
We observe that this simple protocol, although not saturating the bound of the distinguishability, qualitatively behaves in a similar way, and retrieves most of the encoded information for strong disorder.
We note that this protocol can be considered as 
a generalization of a spin-echo protocol with constraints, which can be used to identify and distinguish MBL phase from Anderson localization~\cite{Serbyn2014echoMBL}.

\section{Propagation of mutual information}
\label{sec:mutualinfo}
One of the unique features of  MBL phase is the logarithmically slow growth of entanglement entropy for an initially quenched product state~\cite{bardarson12unbounded,serbyn13slowgrowth,Serbyn_2014_quench}, which has been well explained by development of phase correlations of localized spins~\cite{serbyn13local,huse14phenomenology}.
Since we are interested in the case where the initial state is mixed, we cannot use the entanglement entropy to characterize the MBL phase. Instead,  the appropriate measure of (total) correlations is the mutual information.
For two subsystems $A$ and $B$, the mutual information between the two is given by  $I(A:B)=S(\rho_A)+S(\rho_B)-S(\rho_{AB})$, where $S (\rho_\mu)$ is the entropy of the state $\rho_\mu$.
In the case of a pure state, the mutual information between a region and its complement coincides (up to a factor 2) with the entanglement entropy~\cite{nielsenchuang}. 
When $A$ and $B$ are uncorrelated, i.e., the combined system is in a tensor product of two states $\rho_{AB} = \rho_A \otimes \rho_B$, the mutual information vanishes.
Non-zero mutual information indicates that two subsystems share correlations.

\begin{figure}
\begin{center}
\includegraphics[width=0.9\columnwidth]{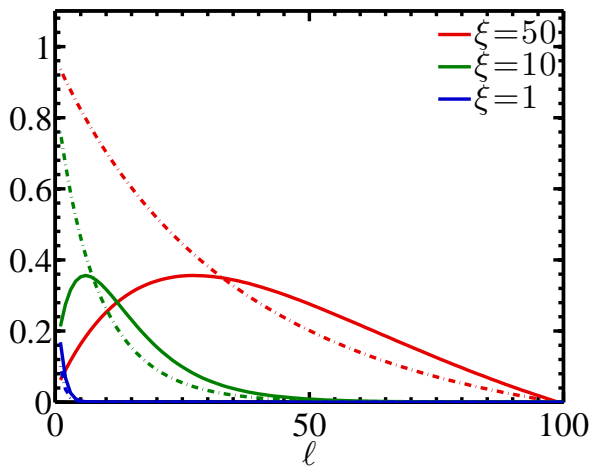}
\caption{Asymptotic value reached by the von Neumann mutual information, in the non-interacting case,
across each cut of a chain of length $N=100$ for initial states $X\pm$ (dash-dotted) or $Z\pm$ (solid lines) for various localization lengths $\xi$.
}
\label{fig:Ixy_vs_V}
\end{center}
\end{figure}

In our case, the initial state in Eq.~\eqref{eq:rhos} exhibits zero correlations with respect to any partitioning of the system.
As the state evolves over time, the initially stored information spreads, 
giving rise to non-trivial correlations across the system.
It is thus interesting to study the propagation of correlations in this setting by analyzing the time evolution of mutual information $I(\ell:N-\ell;t)$ between the first $\ell$ spins and the rest of the chain.
Here, we also consider 
the 2-R\'enyi mutual information, $I_2(A:B)=S_2(\rho_A)+S_2(\rho_B)-S_2(\rho_{AB})$, 
where $S_2(\rho)=-\log[ \mathrm{tr} \rho^2] $ is the R\'enyi entropy of second order.
The 2-R\'enyi entropies can be efficiently computable for large systems, if their
state is given by an MPO, and can be measured from experiments~\cite{Daley:2012bd,Pichler:2013bs,Islam:2015cm}.
For any state of the form \eqref{eq:rhos}, the entropy of the full system, 
which remains constant during the evolution, is $S_2(\rho)=N-1$. 
This provides a bound on the maximum amount of mutual information in our scenario; since the maximum possible entropy for a subsystem corresponds to the number of
spins, the largest allowed value for the mutual information 
is $1$~\footnote{Notice that the maximally mixed states are also the thermal states at infinite temperature,
so that this should be the final value of the mutual information if the system would thermalize
in the case of $X\pm$ initial states.
In our system, total polarization $\sum_i S_z^{[i]}$ is conserved. Thus, the maximum entropy of the initial state $Z\pm$ has to be modified accordingly. These corrections vanish in thermodynamical limit. See Appendix~\ref{app:XY}}.
This is in contrast to the case of entanglement entropy, where the entropy 
grows extensively with the system size due to many-body correlations.
In our case, such situation does not arise as the system starts out in maximally mixed state.

\emph{Non-interacting case.}--- 
Both $I(\ell:N-\ell;t)$ and 
$I_2(\ell:N-\ell;t)$  
can be efficiently computed for any bipartition and time, using the exact time evolution \eqref{eq:rhoZ_spin} 
and \eqref{eq:rhoX_spin}. 
We obtain
\begin{align}
I^{X\pm}(\ell:N-\ell,t)=&\mathrm{H}_b\left(\frac{1+\sqrt{\mathcal{V}_{\ell}}}{2}\right),
\label{eq:Ix_xy}\\
I^{Z\pm}(\ell:N-\ell,t)=&\mathrm{H}_b\left(\frac{\mathcal{V}_{\ell}}{2}\right)+\mathrm{H}_b\left(\frac{1+\mathcal{V}_{\ell}}{2}\right)-1,
\label{eq:Iz_xy}
\end{align}
where $\mathrm{H}_b(p)=-p \log_2(p)-(1-p)\log_2(1-p)$ is the binary entropy function,
and 
 \begin{align}
 I_2^{X\pm}(\ell:N-\ell,t)=&1-\log\left(1+\mathcal{V}_{\ell}\right), 
 \label{eq:I2x_xy}\\
 I_2^{Z\pm}(\ell:N-\ell,t)=&1-\log\left(1+\mathcal{V}_{\ell}^2\right)-\log\left(1+(1-\mathcal{V}_{\ell})^2\right),
 \label{eq:I2z_xy}
 \end{align}
where we introduced the function 
$\mathcal{V}_{\ell}(t)=\sum_{r=0}^{\ell-1}|V_r(t)|^2=1-\sum_{r=\ell}^{N-1}|V_r(t)|^2$, which characterizes the probability that the single excitation remains in the the subsystem of size $\ell$.
We find that, in both cases, the mutual information is only a function of $\mathcal{V}_\ell(t)$.

Initially, $\mathcal{V}_{\ell}(t=0)=1$ for any $\ell$, so that the initial correlations vanish for all partitions. 
This is because, in order to develop non-vanishing mutual information across a certain 
cut $\ell$, the particle must propagate more than $\ell$ sites.
In the localized regime, this probability is exponentially suppressed even in asymptotically long time, e.g. 
$|V_p|^2\propto e^{-p/\xi}$, hence 
$\mathcal{V}_{\ell}$ remains close to zero for a partition $\ell>\xi$.
This implies that
the localization length also determines  the asymptotic behavior of the 
mutual information for each cut, as illustrated in Fig.  \ref{fig:Ixy_vs_V}. 

\emph{Interacting case.}---
 we first estimate the qualitative behavior of the many body localized regime by making use of the $l$-bit 
phenomenological model in Eq.~\eqref{eq:lbits}.
Similar to the previous section, we assume that our system is deep in the localized regime such that the initial state corresponds to a pure state of the first $l$-bit, $(1+\tau_x^{[0]})/2$, 
and a maximally mixed state for the rest of the chain.
Truncating Eq.~\eqref{eq:lbits}
at two-body interactions,
we compute the exact evolution and corresponding time-dependent mutual information of this initial state.
For a bipartition $(\ell:N-\ell)$ we obtain
\begin{align}
I(\ell:N-\ell;t)&=\mathrm{H}_b\left(\frac{1+x(\ell,t)}{2}\right), \\
I_2(\ell:N-\ell;t)&=1-\log(1+x(\ell,t)^2),
\end{align}
In contrast to the non-interacting case, the mutual information across each cut can reach any arbitrary (allowed) value
in a time that grows exponentially with the size of the partition, $\ell$.

\begin{figure}
 \subfigure[ Non-interacting $\rho_{X+}$]{
   \resizebox{.45\columnwidth}{!}{
     \includegraphics{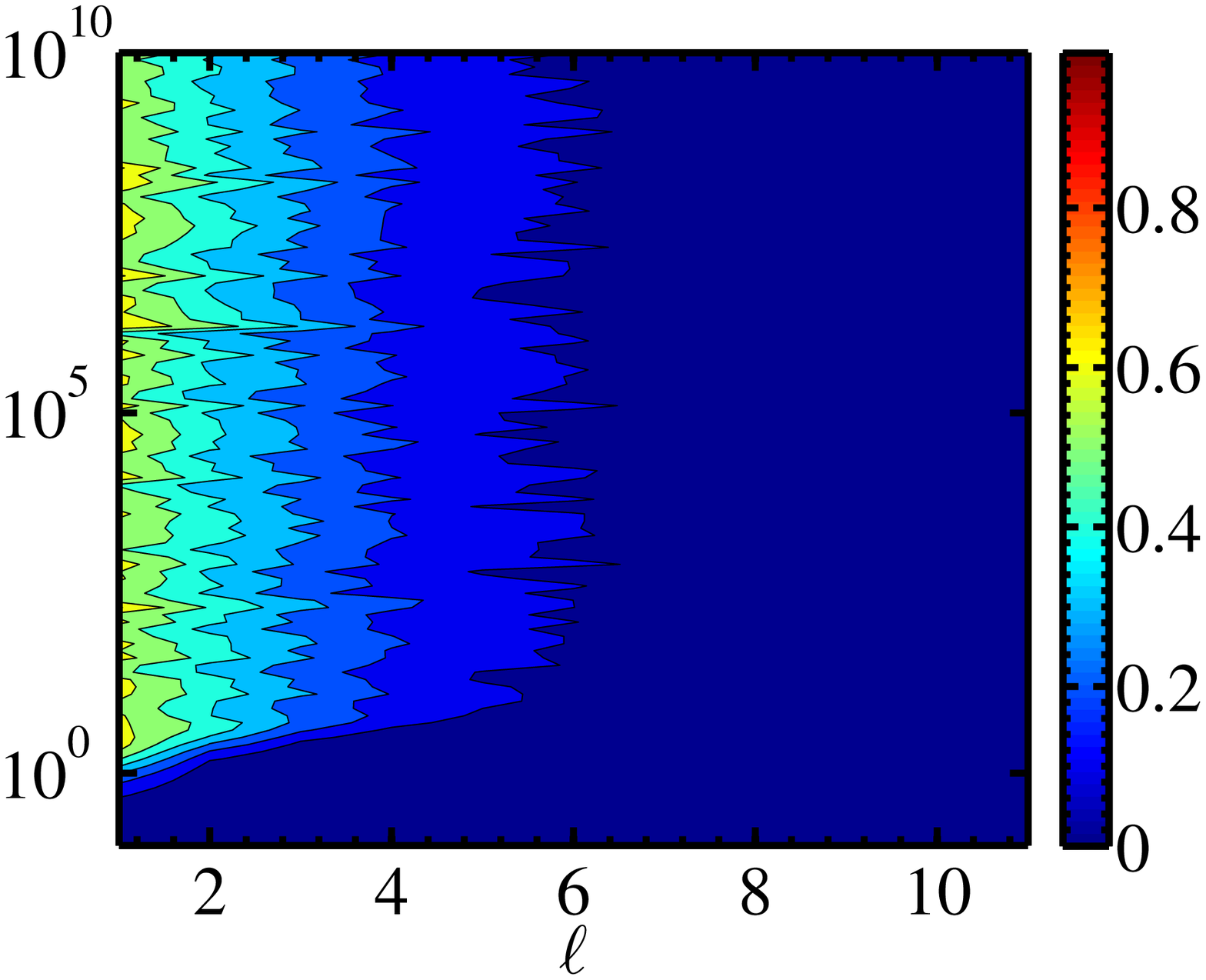}
     }
     }
 \subfigure[ Interacting $\rho_{X+}$]{
   \resizebox{.45\columnwidth}{!}{
     \includegraphics{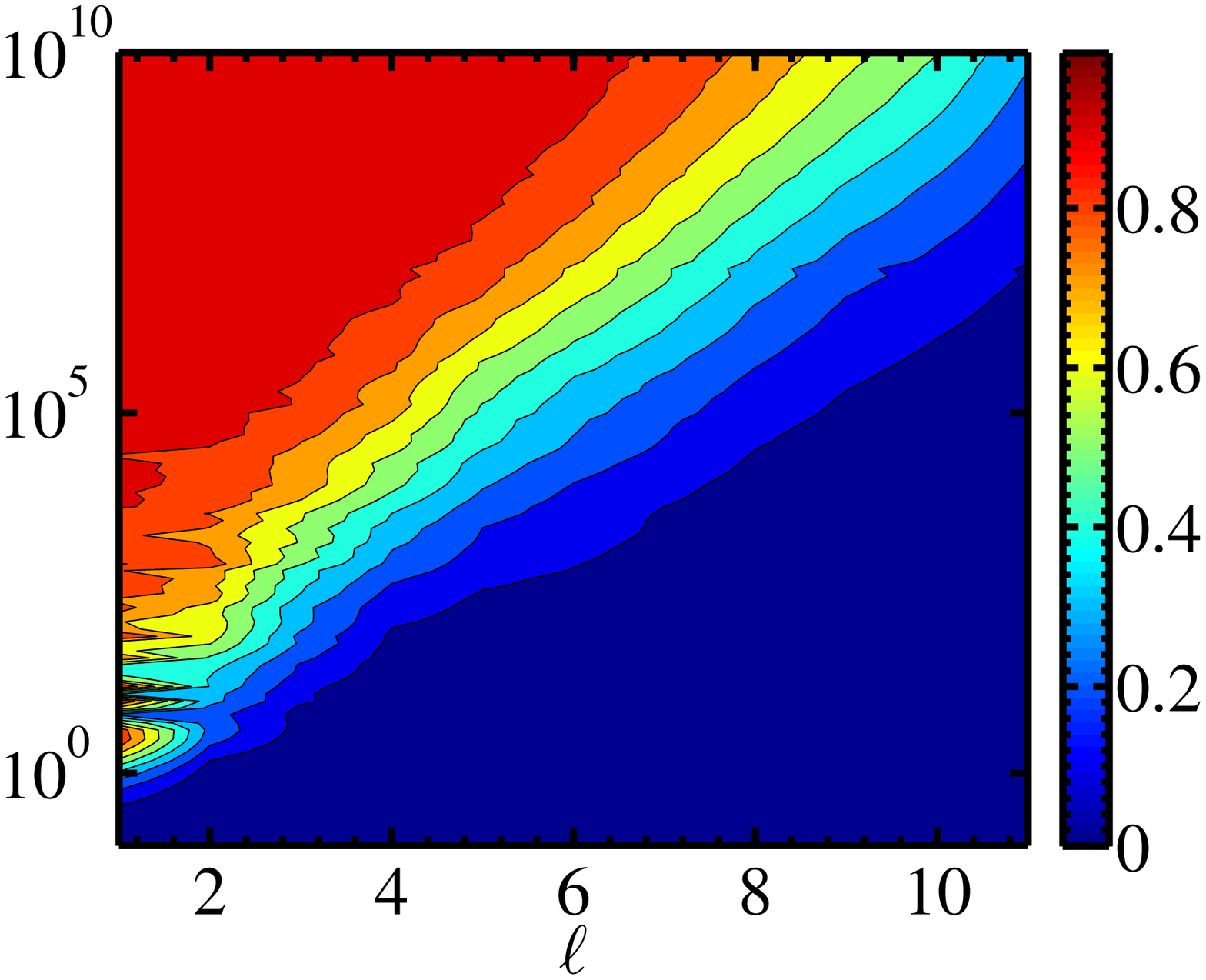}
     }} 
 \subfigure[ Non-interacting $\rho_{Z+}$]{
   \resizebox{.45\columnwidth}{!}{
     \includegraphics{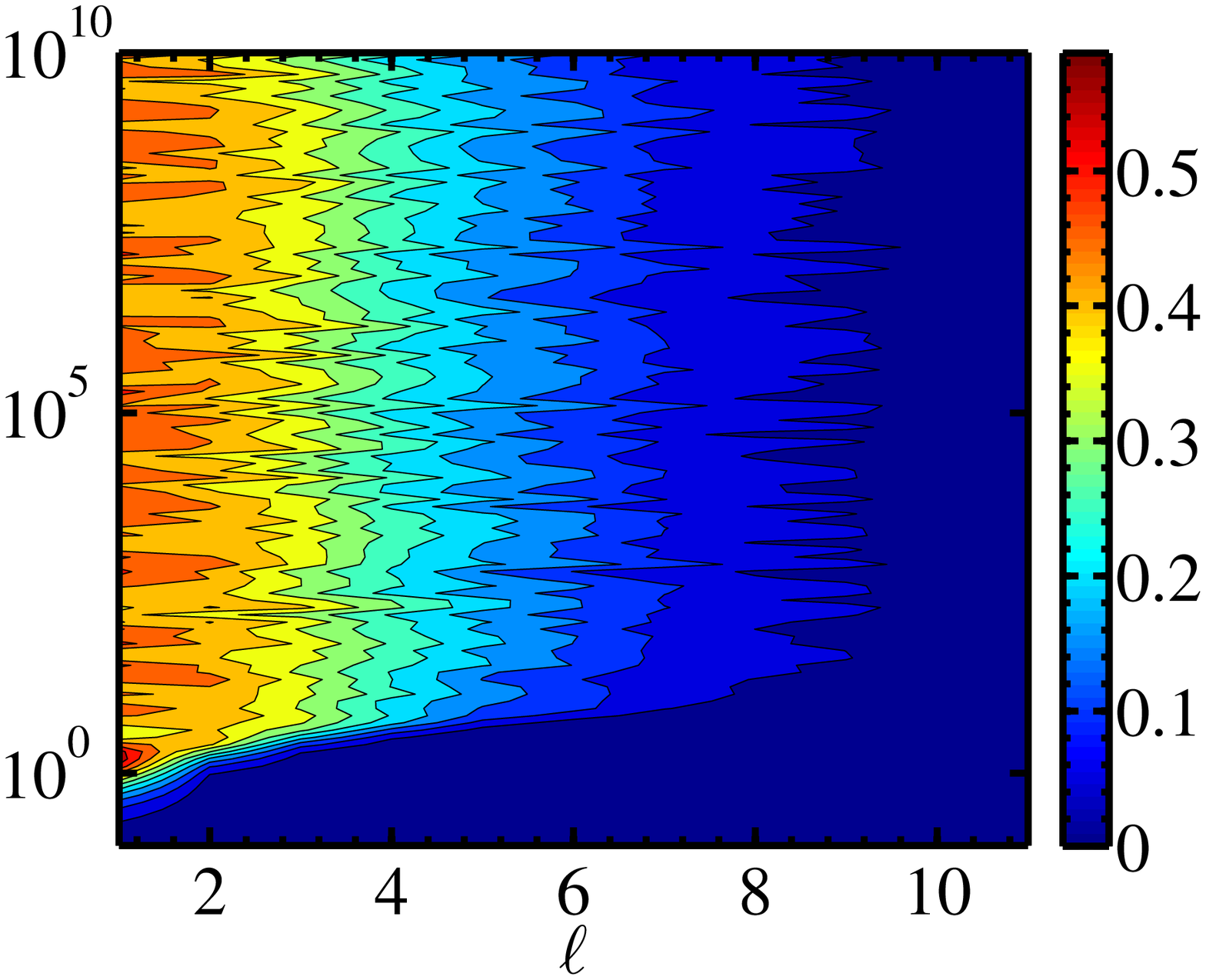}
          }}	
 \subfigure[ Interacting $\rho_{Z+}$]{
   \resizebox{.45\columnwidth}{!}{
     \includegraphics{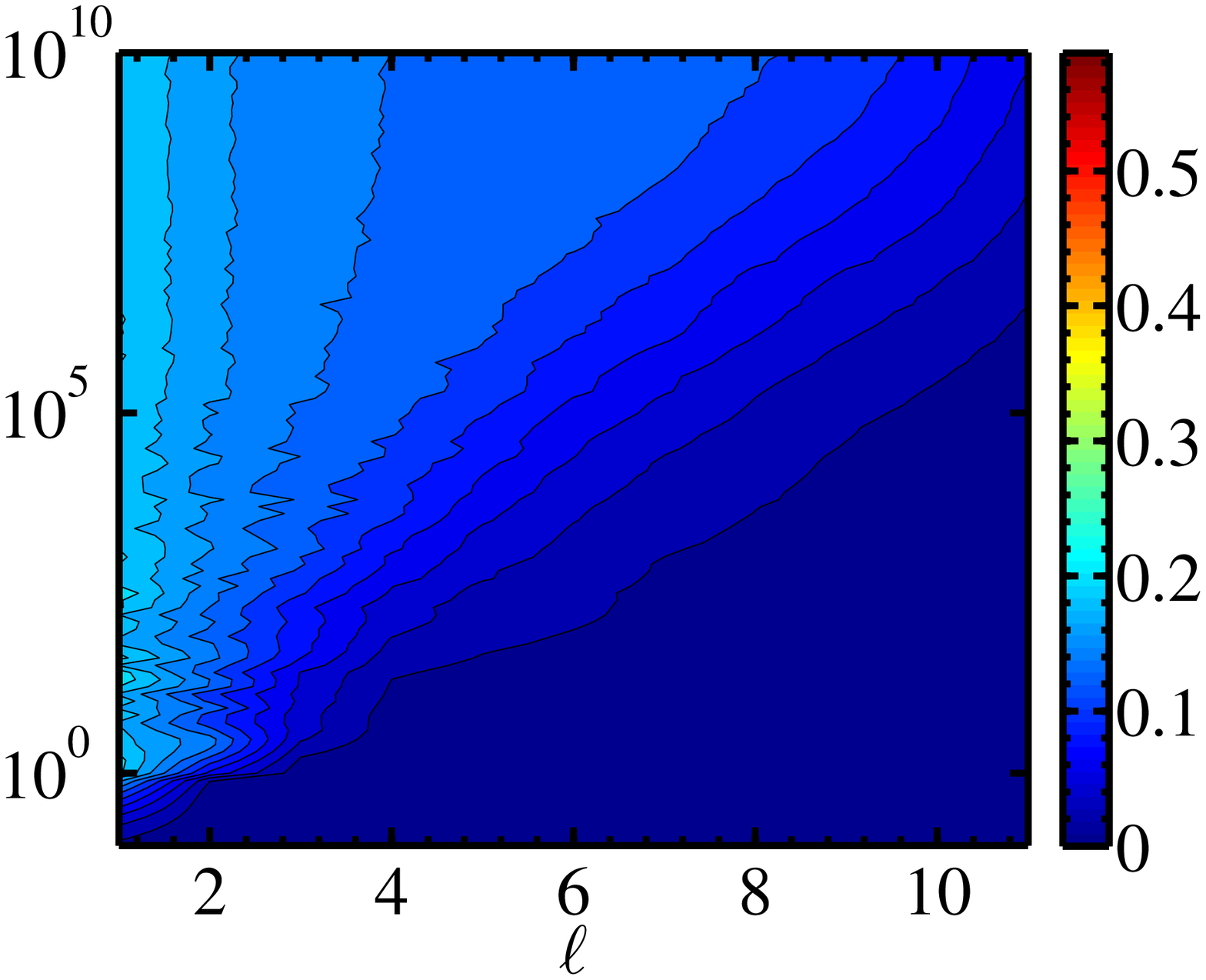}
     }} 
\caption{
Comparison of correlation spreading in the Anderson (left column) and many body localized (right column) scenario, from exact calculations on a chain of size $N=12$,
and averaging over 20 disorder realizations.
The contour plots show the mutual information, $I(\ell:N-\ell,t)$, for initial states  $X+$ (upper row) and $Z+$ (lower row).
In order to visualize the spreading of correlations, we show results for disorder strength $h=1.5$ in the non-interacting case,
while in the interacting case, $h=10$.
}
\label{fig:mutual_info_prop_Alt}
\end{figure}

We compare the simplified calculations above with the exact numerical simulations of the real time dynamics of the considered systems.
The differences between the localized phases in the non-interacting and the interacting cases can be appreciated from 
 Fig.~\ref{fig:mutual_info_prop_Alt}.
 In the former, correlations propagate fast at the beginning, but only to a certain range, and remain 
 localized for arbitrarily long times. The asymptotic values for various cuts follow the 
 qualitative estimation shown in Fig.~\ref{fig:Ixy_vs_V}.
In the interacting case, in contrast, mutual information with respect to every cut   
attains a large value, even though it takes exponentially long in $\ell$. 

\section{Classical Simulability}
\label{sec:errors}

We turn to the simulability of the time dynamics using MPO representations.
Intuitively, one expects that 
a tensor network description of a many-body state is efficient 
when the system is strongly localized.
For the case of pure state evolution, 
this can be quantified by the entanglement entropy; if the entanglement entropy across every partition remains small, a matrix product state provides an efficient description of the system.
Since an MBL system exhibits unbounded growth of entanglement entropy, the required bond dimension to accurately simulate the dynamics grows at least polynomially in time~\cite{znidaric08xxz}.
In the mixed state scenario, the entropy of a certain partition is not the proper figure of merit.
For instance, the maximally mixed state, $\rho\propto \Id^{\otimes L}$, for which any reduced state has maximal entropy, 
has an exact MPO expression with bond dimension $D=1$. 
A better quantifier of the representability as a MPO is the truncation error when using a fixed bond dimension.
This error is defined as the distance between the reference state and the MPO truncated approximation.
Since we do not have access to the exact evolved state, here 
we take as reference the simulated state with the maximum bond dimension $D=80$, and compute the error induced by a smaller value, $D_{\mathrm{cut}}$.
More specifically, we compute the Euclidean distance between the vectorized MPOs, $\epsilon=\|\rho_D-\rho_{D_{\mathrm{cut}}}\|_2$, 
which in the following we simply call \emph{truncation error}.
We note that this truncation error occurs exclusively in the interacting case. For the non-interacting model, $\alpha =0$,
as shown in the previous section,
the time evolution of initial states of the form (\ref{eq:rhos}) is exactly given at any time by a MPO with very 
small bond dimension, irrespective of the disorder strength.

\begin{figure}
\begin{center}
\includegraphics[width=0.9\columnwidth]{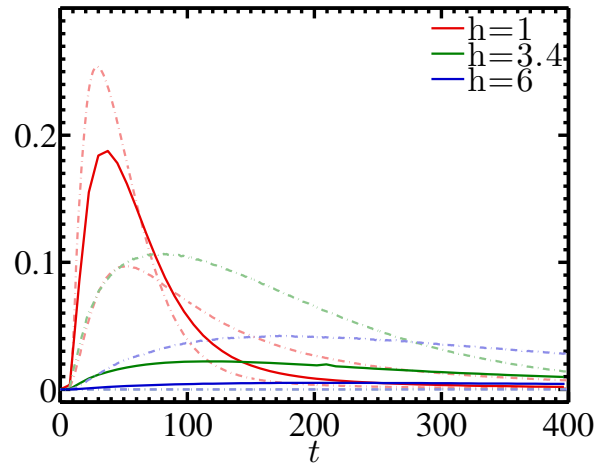}
\caption{Variation of the truncation error evolution for different disorder strengths, in a chain of size $L=40$ and initial state $\rho_{Z+}$.
We show the average truncation error (solid lines) corresponding to a cut with $D_{\mathrm{cut}}=60$ with respect to $D=80$ as well as two instances of disordered potentials $h_i$ that exhibit extremely different behavior (dot-dashed lines). For a weak disorder $h=1$ (red), the truncation error reaches its maximum at short time, followed by rapid decreases. Such a non-monotonic behavior is not clear (or sometimes absent) in the case of strong disorder.
}
\label{fig:truncation_n2-9}
\end{center}
\end{figure}

We simulate the evolution of initial states $\rho_{\varphi}$ in Eq.~\eqref{eq:rhos} for $|\varphi\rangle=|X\pm\rangle$, $|Z\pm\rangle$, for 
chains of lengths $L=40$, and for various disorder strengths.
In each case, we compute the truncation error $\epsilon$ along the evolution for $D_{\mathrm{cut}}=60$, until a maximum time $t=400$.
We observe that the maximum truncation error generally decreases as we 
increase the disorder strength, $h$ (Fig. \ref{fig:truncation_n2-9}). 
For small $h$, the error grows fast at the beginning, and peaks at short times $t\approx 10-50$, to drop afterwards (with 
small fast oscillations that do not alter the overall tendency).
We denote the time at which the truncation error reaches its maximum as $t_\textrm{max}$ to characterize the time scale of simulability.
This non-monotonic behavior softens as the disorder gets stronger, and for the largest values of $h$, we cannot observe a maximum 
in the error within our time window, $t\in[0,\ 400]$. 
Instead, the truncation error seems to increase over the entire time evolution, although 
at much slower rate.

\begin{figure}
\begin{center}
\includegraphics[width=0.9\columnwidth]{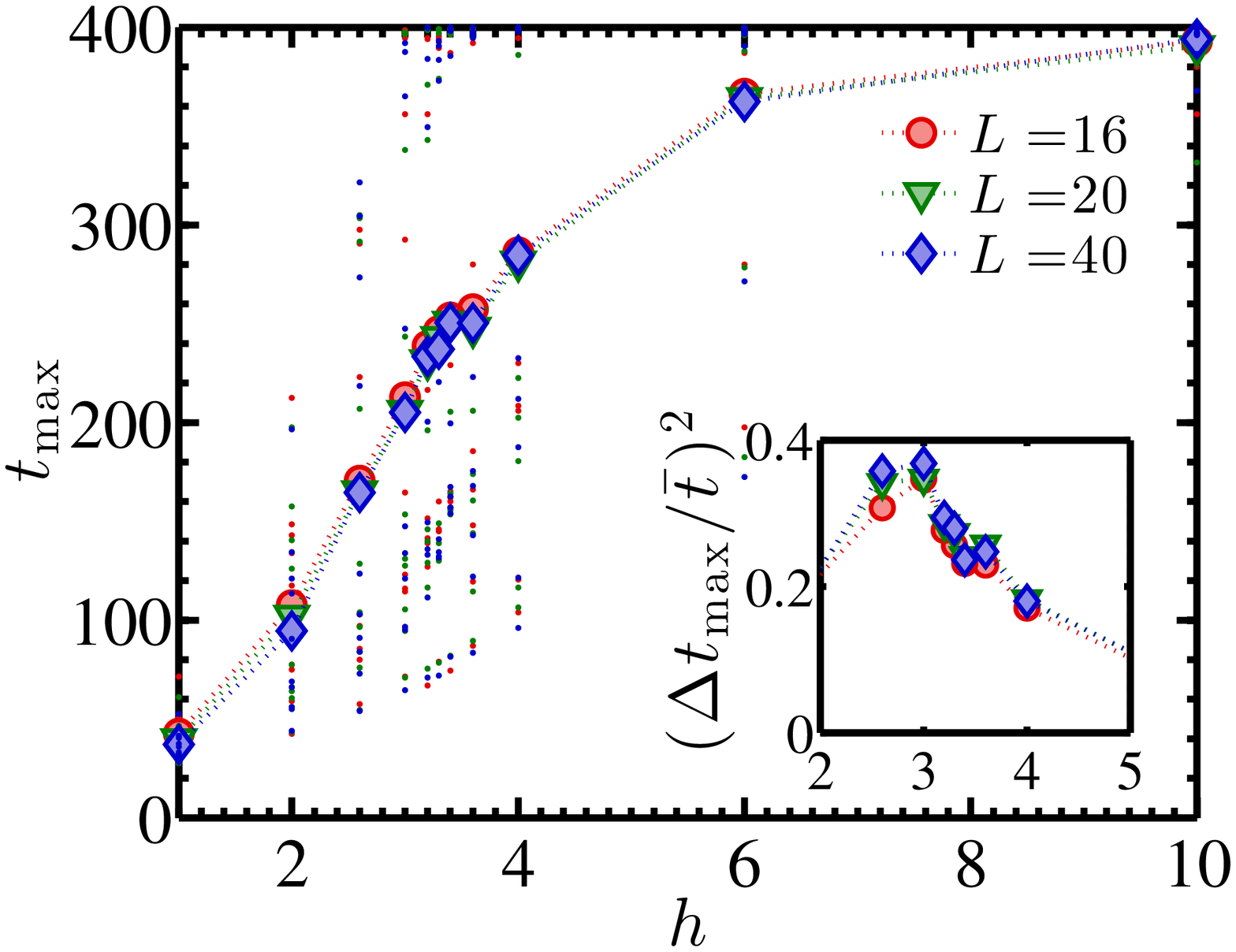}\\
\includegraphics[width=0.9\columnwidth]{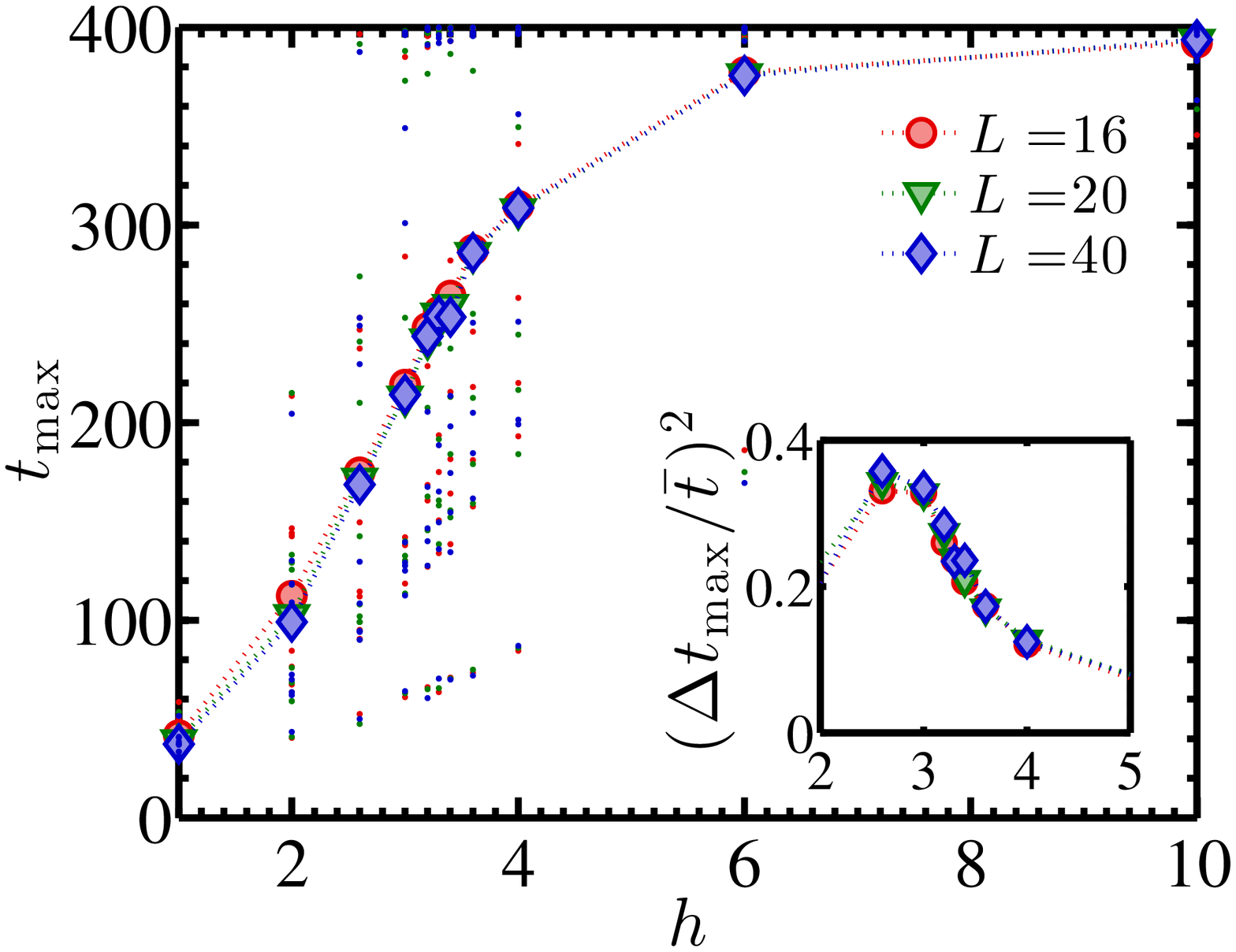}\\
\caption{Average over 10 instances of the time $t_{\mathrm{max}}$ at which the maximum truncation error is attained, for initial states $\rho_{Z+}$ (above) and  $\rho_{X+}$ (below) for system sizes $L=16$ (blue circles), $20$ (green triangles) and $40$ (blue diamonds). The scattered points, in the color corresponding to system size, show the individual location of the maximum for each instance. 
The inset shows the (normalized) variance of the same quantity over this sample, which exhibits a peak near the critical $h_c$.}
\label{fig:truncation_avg}
\end{center}
\end{figure}

Once the truncation error reaches a certain threshold, the MPO should not be taken as a faithful description of the evolved state.
Nevertheless, the dynamical behavior of the error itself provides information about correlations being developed in the system 
and is thus a non-local probe. Therefore one might expect that $t_\textrm{max}$ can diagnose the presence or absence of localization.
To this end, we further investigate the behavior of $t_{\mathrm{max}}$ for each instance of disordered potentials, as summarized in Fig.~\ref{fig:truncation_avg}.
As a generic trend, we observe that $t_\textrm{max}$ tends to grow with $h$, at a rate that seems to increase for larger values of the disorder strength,
although for some instance we observe some non-monotonic behavior at intermediate values of $h$.
The less \emph{localizing} an instance seems, the later this larger slope can be appreciated, with the instances in Fig. \ref{fig:truncation_n2-9} 
being again the extreme cases.

If we average over a sample of $10$ instances, the average $t_{\mathrm{tmax}}$ does not show this increasing rate (Fig.~\ref{fig:truncation_avg}).
Interestingly, the (normalized) variance of $t_{\mathrm{max}}$ over the sample (inset) exhibits a clear peak in the 
region of disorder strengths $h\approx 3$, close to ergodic-MBL phase transition point~\cite{pal2010,luitz2015}.
This result suggests that the simulability itself of the dynamics with MPO may be an indicator of the MBL transition.
Indeed, it has been predicted that, 
when approaching the phase transition from the thermal side, the system may cross through a
Griffiths phase, where rare (quasi)-localized regions govern dynamical observables~\cite{agarwal15grif,vosk15prx}.
 Large fluctuations in $t_\textrm{max}$ are consistent with such expectations.

\section{Conclusion}
\label{sec:conclu}
We have discussed several information theoretical aspects of localized phases.
Using MPO representations, we investigate the infinite temperature dynamics of information that is initially encoded in a single qubit.
We quantify the amount of information that remains near the vicinity of the initial encoding position by using the distinguishability of many-body density matrices.
Moreover, we explore the propagation dynamics of mutual information, which exhibit qualitatively distinct behaviors in thermalizing, Anderson localized and MBL systems.
Unlike the entanglement entropy of a pure state under quench dynamics, 
the mutual information remains upper bounded in all cases.

Finally, we demonstrate that localized dynamics are reflected in the classical simulability of
MBL time evolution by estimating the truncation error of their approximate MPO description.
We have observed that this error, which can be interpreted as a non-local probe of correlations, may qualitatively capture the location of the localization transition. 
In particular, the variance of the time at which the simulation error reaches its maximum value (over different disorder realizations), exhibits signatures of peaking near the nominal critical point~\cite{pal2010,luitz2015,Singh_numerics_2015}.  

\section{Acknowledgements}
\label{sec:acknow}
It is a pleasure to gratefully acknowledge the insights of and discussions with E. Altman, D. Huse and M. Knap. 
Part of this work was developed while some of the authors visited the KITP Santa Barbara, supported partially by the National Science Foundation under Grant No. NSF PHY-1125915. This work was also supported in part by the NSF PHY-1654740, CUA, the Miller Institute for Basic Research in Science, AFOSR MURI and the Moore Foundation.

\newpage

\appendix
\section{Exact results for the non-interacting chain}
\label{app:XY}
\emph{Exact MPO for the time evolved states.}---
The time evolution of initial states of the form \eqref{eq:rhos} can be computed using the
exact solution of the non-interacting XY chain,
and the resulting time dependent density operators are MPO~\cite{mcculloch2007,pirvu10mpo} of constant bond dimension.
Here we show explicitly the tensors for such decompositions.
Specifically, the $\rho_{X\pm}(t)$ states \eqref{eq:rhoX_spin} correspond to a MPO of bond dimension 2, specified by the operator valued matrices
\begin{align}
&M_{X\pm}^{(0)}(t)=\frac{1}{2}\left (
\begin{array}{cc}
\pm \sigma_n^z &\Id
\end{array}
\right ), \nn \\
&M_{X\pm}^{(n)}(t)=\frac{1}{2}\left (
\begin{array}{cc}
\sigma_n^z&
V_n \sigma_n^-+V_n^* \sigma_n^+\\
0&\Id
\end{array}
\right ),  \, 0< n<L-1, \nn \\
&M_{X\pm}^{(L-1)}(t)=\frac{1}{2}\left (
\begin{array}{c}
V_n \sigma_n^-+V_n^* \sigma_n^+
\\ \Id
\end{array}
\right ).
\end{align}
Correspondingly, for the time evolved $\rho_{Z\pm}(t)$  states \eqref{eq:rhoZ_spin}, the MPO is given by
the following tensors
\begin{align}
&M_{Z\pm}^{(0)}(t)=\frac{1}{2}\left (
\begin{array}{cccc}
\Id &
\mp V_n \sigma_n^- &
\mp V_n^* \sigma_n^+ &
|V_n|^2 (\Id\pm\sigma_n^z)
\end{array}
\right ), \nn \\
&M_{Z\pm}^{(0<n<L-1)}(t)=\frac{1}{2}\left (
\begin{array}{cccc}
\Id &
\mp V_n \sigma_n^- &
\mp V_n^* \sigma_n^+ &
|V_n|^2 (\Id\pm\sigma_n^z) \\
0& \sigma_n^z & 0 & 2V_n^* \sigma_n^+ \\
0&0&\sigma_n^z &2V_n \sigma_n^- \\
0&0&0&\Id
\end{array}
\right ),  \nn \\
&M_{Z\pm}^{(L-1)}(t)=\frac{1}{2}\left (
\begin{array}{c}
|V_n|^2 (\Id\pm\sigma_n^z) \\
2V_n^* \sigma_n^+ \\
2V_n \sigma_n^- \\
\Id
\end{array}
\right ).
\end{align}
\emph{Thermal equilibrium states.}---
Total polarization, $\sum_iS_i^{z}$, or in the fermionic language, 
total number of particles, $\hat{N}=\sum_p c_p^{\dagger}c_p=\sum_k b_k^{\dagger}b_k$, is conserved in the system.
Thus, equilibration is only possible to a generalized Gibbs ensemble (GGE) compatible with the 
initial energy densities and number of particles, of the form
$\rho_{\mathrm{GGE}}(\beta,\mu)\propto \exp({-\beta H -\mu \hat{N}})$.

In terms of the diagonal modes,
\begin{align}
\rho_{\mathrm{GGE}}=\prod_k \frac{\Id+(e^{-\beta \Lambda_k-\mu}-1)b_k^{\dagger}b_k}{1+e^{-\beta \Lambda_k-\mu}},
\end{align}
which has energy 
$E_{\mathrm{GGE}}(\beta,\mu)=H_0+\sum_k \frac{\Lambda_k e^{-\beta \Lambda_k-\mu}}{1+e^{-\beta \Lambda_k-\mu}}$
and number of particles
$N_{\mathrm{GGE}}(\beta,\mu)=\sum_k \frac{e^{-\beta \Lambda_k-\mu}}{1+e^{-\beta \Lambda_k-\mu}}.$

The initial states $\rho_{X\pm}$ correspond thus to  $\beta=\mu=0$, with $E_{\mathrm{GGE}}(0,0)=0$
and $N_{\mathrm{GGE}}(0,0)=L/2$, so the upper bound of the mutual information $1$ really corresponds to thermalization.
For $\rho_{Z\pm}$ states, instead, the energy is $\pm h_0/2$ and the number of particles $(L\pm1)/2$.
The values of $\beta$ and $\mu$ that produce the GGE with the same 
conserved quantities can be determined numerically.

\bibliographystyle{apsrev4-1}
\bibliography{MBL_QI}

\begin{thebibliography}{49}%
\makeatletter
\providecommand \@ifxundefined [1]{%
 \@ifx{#1\undefined}
}%
\providecommand \@ifnum [1]{%
 \ifnum #1\expandafter \@firstoftwo
 \else \expandafter \@secondoftwo
 \fi
}%
\providecommand \@ifx [1]{%
 \ifx #1\expandafter \@firstoftwo
 \else \expandafter \@secondoftwo
 \fi
}%
\providecommand \natexlab [1]{#1}%
\providecommand \enquote  [1]{``#1''}%
\providecommand \bibnamefont  [1]{#1}%
\providecommand \bibfnamefont [1]{#1}%
\providecommand \citenamefont [1]{#1}%
\providecommand \href@noop [0]{\@secondoftwo}%
\providecommand \href [0]{\begingroup \@sanitize@url \@href}%
\providecommand \@href[1]{\@@startlink{#1}\@@href}%
\providecommand \@@href[1]{\endgroup#1\@@endlink}%
\providecommand \@sanitize@url [0]{\catcode `\\12\catcode `\$12\catcode
  `\&12\catcode `\#12\catcode `\^12\catcode `\_12\catcode `\%12\relax}%
\providecommand \@@startlink[1]{}%
\providecommand \@@endlink[0]{}%
\providecommand \url  [0]{\begingroup\@sanitize@url \@url }%
\providecommand \@url [1]{\endgroup\@href {#1}{\urlprefix }}%
\providecommand \urlprefix  [0]{URL }%
\providecommand \Eprint [0]{\href }%
\providecommand \doibase [0]{http://dx.doi.org/}%
\providecommand \selectlanguage [0]{\@gobble}%
\providecommand \bibinfo  [0]{\@secondoftwo}%
\providecommand \bibfield  [0]{\@secondoftwo}%
\providecommand \translation [1]{[#1]}%
\providecommand \BibitemOpen [0]{}%
\providecommand \bibitemStop [0]{}%
\providecommand \bibitemNoStop [0]{.\EOS\space}%
\providecommand \EOS [0]{\spacefactor3000\relax}%
\providecommand \BibitemShut  [1]{\csname bibitem#1\endcsname}%
\let\auto@bib@innerbib\@empty
\bibitem [{\citenamefont {Deutsch}(1991)}]{Deutsch_1991_ETH}%
  \BibitemOpen
  \bibfield  {author} {\bibinfo {author} {\bibfnamefont {J.~M.}\ \bibnamefont
  {Deutsch}},\ }\href {\doibase 10.1103/PhysRevA.43.2046} {\bibfield  {journal}
  {\bibinfo  {journal} {Phys. Rev. A}\ }\textbf {\bibinfo {volume} {43}},\
  \bibinfo {pages} {2046} (\bibinfo {year} {1991})}\BibitemShut {NoStop}%
\bibitem [{\citenamefont {Srednicki}(1994)}]{Srednicki_1994_ETH}%
  \BibitemOpen
  \bibfield  {author} {\bibinfo {author} {\bibfnamefont {M.}~\bibnamefont
  {Srednicki}},\ }\href {\doibase 10.1103/PhysRevE.50.888} {\bibfield
  {journal} {\bibinfo  {journal} {Phys. Rev. E}\ }\textbf {\bibinfo {volume}
  {50}},\ \bibinfo {pages} {888} (\bibinfo {year} {1994})}\BibitemShut
  {NoStop}%
\bibitem [{\citenamefont {{Anderson, P W}}(1958)}]{Anderson:2011wp}%
  \BibitemOpen
  \bibfield  {author} {\bibinfo {author} {\bibnamefont {{Anderson, P W}}},\
  }\href@noop {} {\bibfield  {journal} {\bibinfo  {journal} {Physical Review}\
  }\textbf {\bibinfo {volume} {109}},\ \bibinfo {pages} {1492} (\bibinfo {year}
  {1958})}\BibitemShut {NoStop}%
\bibitem [{\citenamefont {Fleishman}\ and\ \citenamefont
  {Anderson}(1980)}]{PhysRevB.21.2366_1980}%
  \BibitemOpen
  \bibfield  {author} {\bibinfo {author} {\bibfnamefont {L.}~\bibnamefont
  {Fleishman}}\ and\ \bibinfo {author} {\bibfnamefont {P.~W.}\ \bibnamefont
  {Anderson}},\ }\href {\doibase 10.1103/PhysRevB.21.2366} {\bibfield
  {journal} {\bibinfo  {journal} {Phys. Rev. B}\ }\textbf {\bibinfo {volume}
  {21}},\ \bibinfo {pages} {2366} (\bibinfo {year} {1980})}\BibitemShut
  {NoStop}%
\bibitem [{\citenamefont {Basko}\ \emph {et~al.}(2006)\citenamefont {Basko},
  \citenamefont {Aleiner},\ and\ \citenamefont {Altshuler}}]{basko2006}%
  \BibitemOpen
  \bibfield  {author} {\bibinfo {author} {\bibfnamefont {D.~M.}\ \bibnamefont
  {Basko}}, \bibinfo {author} {\bibfnamefont {I.~L.}\ \bibnamefont {Aleiner}},
  \ and\ \bibinfo {author} {\bibfnamefont {B.~L.}\ \bibnamefont {Altshuler}},\
  }\href {\doibase http://dx.doi.org/10.1016/j.aop.2005.11.014} {\bibfield
  {journal} {\bibinfo  {journal} {Annals of Physics}\ }\textbf {\bibinfo
  {volume} {321}},\ \bibinfo {pages} {1126 } (\bibinfo {year}
  {2006})}\BibitemShut {NoStop}%
\bibitem [{\citenamefont {Oganesyan}\ and\ \citenamefont
  {Huse}(2007)}]{PhysRevB.75.155111_2007}%
  \BibitemOpen
  \bibfield  {author} {\bibinfo {author} {\bibfnamefont {V.}~\bibnamefont
  {Oganesyan}}\ and\ \bibinfo {author} {\bibfnamefont {D.~A.}\ \bibnamefont
  {Huse}},\ }\href {\doibase 10.1103/PhysRevB.75.155111} {\bibfield  {journal}
  {\bibinfo  {journal} {Phys. Rev. B}\ }\textbf {\bibinfo {volume} {75}},\
  \bibinfo {pages} {155111} (\bibinfo {year} {2007})}\BibitemShut {NoStop}%
\bibitem [{\citenamefont {\v{Z}nidari\v{c}}\ \emph {et~al.}(2008)\citenamefont
  {\v{Z}nidari\v{c}}, \citenamefont {Prosen},\ and\ \citenamefont
  {Prelov\v{s}ek}}]{znidaric08xxz}%
  \BibitemOpen
  \bibfield  {author} {\bibinfo {author} {\bibfnamefont {M.}~\bibnamefont
  {\v{Z}nidari\v{c}}}, \bibinfo {author} {\bibfnamefont {T.}~\bibnamefont
  {Prosen}}, \ and\ \bibinfo {author} {\bibfnamefont {P.}~\bibnamefont
  {Prelov\v{s}ek}},\ }\href {\doibase 10.1103/PhysRevB.77.064426} {\bibfield
  {journal} {\bibinfo  {journal} {Phys. Rev. B}\ }\textbf {\bibinfo {volume}
  {77}},\ \bibinfo {pages} {064426} (\bibinfo {year} {2008})}\BibitemShut
  {NoStop}%
\bibitem [{\citenamefont {Pal}\ and\ \citenamefont {Huse}(2010)}]{pal2010}%
  \BibitemOpen
  \bibfield  {author} {\bibinfo {author} {\bibfnamefont {A.}~\bibnamefont
  {Pal}}\ and\ \bibinfo {author} {\bibfnamefont {D.~A.}\ \bibnamefont {Huse}},\
  }\href {\doibase 10.1103/PhysRevB.82.174411} {\bibfield  {journal} {\bibinfo
  {journal} {Phys. Rev. B}\ }\textbf {\bibinfo {volume} {82}},\ \bibinfo
  {pages} {174411} (\bibinfo {year} {2010})}\BibitemShut {NoStop}%
\bibitem [{\citenamefont {Bardarson}\ \emph {et~al.}(2012)\citenamefont
  {Bardarson}, \citenamefont {Pollmann},\ and\ \citenamefont
  {Moore}}]{bardarson12unbounded}%
  \BibitemOpen
  \bibfield  {author} {\bibinfo {author} {\bibfnamefont {J.~H.}\ \bibnamefont
  {Bardarson}}, \bibinfo {author} {\bibfnamefont {F.}~\bibnamefont {Pollmann}},
  \ and\ \bibinfo {author} {\bibfnamefont {J.~E.}\ \bibnamefont {Moore}},\
  }\href {\doibase 10.1103/PhysRevLett.109.017202} {\bibfield  {journal}
  {\bibinfo  {journal} {Phys. Rev. Lett.}\ }\textbf {\bibinfo {volume} {109}},\
  \bibinfo {pages} {017202} (\bibinfo {year} {2012})}\BibitemShut {NoStop}%
\bibitem [{\citenamefont {Serbyn}\ \emph
  {et~al.}(2013{\natexlab{a}})\citenamefont {Serbyn}, \citenamefont
  {Papi{\'{c}}},\ and\ \citenamefont {Abanin}}]{serbyn13slowgrowth}%
  \BibitemOpen
  \bibfield  {author} {\bibinfo {author} {\bibfnamefont {M.}~\bibnamefont
  {Serbyn}}, \bibinfo {author} {\bibfnamefont {Z.}~\bibnamefont {Papi{\'{c}}}},
  \ and\ \bibinfo {author} {\bibfnamefont {D.~A.}\ \bibnamefont {Abanin}},\
  }\href {\doibase 10.1103/PhysRevLett.110.260601} {\bibfield  {journal}
  {\bibinfo  {journal} {Phys. Rev. Lett.}\ }\textbf {\bibinfo {volume} {110}},\
  \bibinfo {pages} {260601} (\bibinfo {year} {2013}{\natexlab{a}})}\BibitemShut
  {NoStop}%
\bibitem [{\citenamefont {Serbyn}\ \emph
  {et~al.}(2013{\natexlab{b}})\citenamefont {Serbyn}, \citenamefont
  {Papi{\'{c}}},\ and\ \citenamefont {Abanin}}]{serbyn13local}%
  \BibitemOpen
  \bibfield  {author} {\bibinfo {author} {\bibfnamefont {M.}~\bibnamefont
  {Serbyn}}, \bibinfo {author} {\bibfnamefont {Z.}~\bibnamefont {Papi{\'{c}}}},
  \ and\ \bibinfo {author} {\bibfnamefont {D.~A.}\ \bibnamefont {Abanin}},\
  }\href {\doibase 10.1103/PhysRevLett.111.127201} {\bibfield  {journal}
  {\bibinfo  {journal} {Phys. Rev. Lett.}\ }\textbf {\bibinfo {volume} {111}},\
  \bibinfo {pages} {127201} (\bibinfo {year} {2013}{\natexlab{b}})}\BibitemShut
  {NoStop}%
\bibitem [{\citenamefont {Serbyn}\ \emph
  {et~al.}(2014{\natexlab{a}})\citenamefont {Serbyn}, \citenamefont {Knap},
  \citenamefont {Gopalakrishnan}, \citenamefont {Papi\ifmmode~\acute{c}\else
  \'{c}\fi{}}, \citenamefont {Yao}, \citenamefont {Laumann}, \citenamefont
  {Abanin}, \citenamefont {Lukin},\ and\ \citenamefont
  {Demler}}]{Serbyn2014echoMBL}%
  \BibitemOpen
  \bibfield  {author} {\bibinfo {author} {\bibfnamefont {M.}~\bibnamefont
  {Serbyn}}, \bibinfo {author} {\bibfnamefont {M.}~\bibnamefont {Knap}},
  \bibinfo {author} {\bibfnamefont {S.}~\bibnamefont {Gopalakrishnan}},
  \bibinfo {author} {\bibfnamefont {Z.}~\bibnamefont
  {Papi\ifmmode~\acute{c}\else \'{c}\fi{}}}, \bibinfo {author} {\bibfnamefont
  {N.~Y.}\ \bibnamefont {Yao}}, \bibinfo {author} {\bibfnamefont {C.~R.}\
  \bibnamefont {Laumann}}, \bibinfo {author} {\bibfnamefont {D.~A.}\
  \bibnamefont {Abanin}}, \bibinfo {author} {\bibfnamefont {M.~D.}\
  \bibnamefont {Lukin}}, \ and\ \bibinfo {author} {\bibfnamefont {E.~A.}\
  \bibnamefont {Demler}},\ }\href {\doibase 10.1103/PhysRevLett.113.147204}
  {\bibfield  {journal} {\bibinfo  {journal} {Phys. Rev. Lett.}\ }\textbf
  {\bibinfo {volume} {113}},\ \bibinfo {pages} {147204} (\bibinfo {year}
  {2014}{\natexlab{a}})}\BibitemShut {NoStop}%
\bibitem [{\citenamefont {Serbyn}\ \emph
  {et~al.}(2014{\natexlab{b}})\citenamefont {Serbyn}, \citenamefont
  {Papi{\'{c}}},\ and\ \citenamefont {Abanin}}]{Serbyn_2014_quench}%
  \BibitemOpen
  \bibfield  {author} {\bibinfo {author} {\bibfnamefont {M.}~\bibnamefont
  {Serbyn}}, \bibinfo {author} {\bibfnamefont {Z.}~\bibnamefont {Papi{\'{c}}}},
  \ and\ \bibinfo {author} {\bibfnamefont {D.~A.}\ \bibnamefont {Abanin}},\
  }\href {\doibase 10.1103/PhysRevB.90.174302} {\bibfield  {journal} {\bibinfo
  {journal} {Phys. Rev. B}\ }\textbf {\bibinfo {volume} {90}},\ \bibinfo
  {pages} {174302} (\bibinfo {year} {2014}{\natexlab{b}})}\BibitemShut
  {NoStop}%
\bibitem [{\citenamefont {{Choi}}\ \emph {et~al.}(2015)\citenamefont {{Choi}},
  \citenamefont {{Yao}}, \citenamefont {{Gopalakrishnan}},\ and\ \citenamefont
  {{Lukin}}}]{choi2015qcmbl}%
  \BibitemOpen
  \bibfield  {author} {\bibinfo {author} {\bibfnamefont {S.}~\bibnamefont
  {{Choi}}}, \bibinfo {author} {\bibfnamefont {N.~Y.}\ \bibnamefont {{Yao}}},
  \bibinfo {author} {\bibfnamefont {S.}~\bibnamefont {{Gopalakrishnan}}}, \
  and\ \bibinfo {author} {\bibfnamefont {M.~D.}\ \bibnamefont {{Lukin}}},\
  }\href@noop {} {\bibfield  {journal} {\bibinfo  {journal} {ArXiv e-prints}\ }
  (\bibinfo {year} {2015})},\ \Eprint {http://arxiv.org/abs/1508.06992}
  {arXiv:1508.06992 [quant-ph]} \BibitemShut {NoStop}%
\bibitem [{\citenamefont {{Yao}}\ \emph {et~al.}(2015)\citenamefont {{Yao}},
  \citenamefont {{Laumann}},\ and\ \citenamefont
  {{Vishwanath}}}]{Yao2015MBL_state_transfer}%
  \BibitemOpen
  \bibfield  {author} {\bibinfo {author} {\bibfnamefont {N.~Y.}\ \bibnamefont
  {{Yao}}}, \bibinfo {author} {\bibfnamefont {C.~R.}\ \bibnamefont
  {{Laumann}}}, \ and\ \bibinfo {author} {\bibfnamefont {A.}~\bibnamefont
  {{Vishwanath}}},\ }\href@noop {} {\bibfield  {journal} {\bibinfo  {journal}
  {ArXiv e-prints}\ } (\bibinfo {year} {2015})},\ \Eprint
  {http://arxiv.org/abs/1508.06995} {arXiv:1508.06995 [quant-ph]} \BibitemShut
  {NoStop}%
\bibitem [{\citenamefont {Vosk}\ and\ \citenamefont
  {Altman}(2014)}]{Vosk_2014}%
  \BibitemOpen
  \bibfield  {author} {\bibinfo {author} {\bibfnamefont {R.}~\bibnamefont
  {Vosk}}\ and\ \bibinfo {author} {\bibfnamefont {E.}~\bibnamefont {Altman}},\
  }\href {\doibase 10.1103/PhysRevLett.112.217204} {\bibfield  {journal}
  {\bibinfo  {journal} {Phys. Rev. Lett.}\ }\textbf {\bibinfo {volume} {112}},\
  \bibinfo {pages} {217204} (\bibinfo {year} {2014})}\BibitemShut {NoStop}%
\bibitem [{\citenamefont {Kj{\"a}ll}\ \emph {et~al.}(2014)\citenamefont
  {Kj{\"a}ll}, \citenamefont {Bardarson},\ and\ \citenamefont
  {Pollmann}}]{PhysRevLett.113.107204_2014}%
  \BibitemOpen
  \bibfield  {author} {\bibinfo {author} {\bibfnamefont {J.~A.}\ \bibnamefont
  {Kj{\"a}ll}}, \bibinfo {author} {\bibfnamefont {J.~H.}\ \bibnamefont
  {Bardarson}}, \ and\ \bibinfo {author} {\bibfnamefont {F.}~\bibnamefont
  {Pollmann}},\ }\href {\doibase 10.1103/PhysRevLett.113.107204} {\bibfield
  {journal} {\bibinfo  {journal} {Phys. Rev. Lett.}\ }\textbf {\bibinfo
  {volume} {113}},\ \bibinfo {pages} {107204} (\bibinfo {year}
  {2014})}\BibitemShut {NoStop}%
\bibitem [{\citenamefont {Pekker}\ \emph {et~al.}(2014)\citenamefont {Pekker},
  \citenamefont {Refael}, \citenamefont {Altman}, \citenamefont {Demler},\ and\
  \citenamefont {Oganesyan}}]{PhysRevX.4.011052_2014}%
  \BibitemOpen
  \bibfield  {author} {\bibinfo {author} {\bibfnamefont {D.}~\bibnamefont
  {Pekker}}, \bibinfo {author} {\bibfnamefont {G.}~\bibnamefont {Refael}},
  \bibinfo {author} {\bibfnamefont {E.}~\bibnamefont {Altman}}, \bibinfo
  {author} {\bibfnamefont {E.}~\bibnamefont {Demler}}, \ and\ \bibinfo {author}
  {\bibfnamefont {V.}~\bibnamefont {Oganesyan}},\ }\href {\doibase
  10.1103/PhysRevX.4.011052} {\bibfield  {journal} {\bibinfo  {journal} {Phys.
  Rev. X}\ }\textbf {\bibinfo {volume} {4}},\ \bibinfo {pages} {011052}
  (\bibinfo {year} {2014})}\BibitemShut {NoStop}%
\bibitem [{\citenamefont {Huse}\ \emph {et~al.}(2014)\citenamefont {Huse},
  \citenamefont {Nandkishore},\ and\ \citenamefont
  {Oganesyan}}]{huse14phenomenology}%
  \BibitemOpen
  \bibfield  {author} {\bibinfo {author} {\bibfnamefont {D.~A.}\ \bibnamefont
  {Huse}}, \bibinfo {author} {\bibfnamefont {R.}~\bibnamefont {Nandkishore}}, \
  and\ \bibinfo {author} {\bibfnamefont {V.}~\bibnamefont {Oganesyan}},\ }\href
  {\doibase 10.1103/PhysRevB.90.174202} {\bibfield  {journal} {\bibinfo
  {journal} {Phys. Rev. B}\ }\textbf {\bibinfo {volume} {90}},\ \bibinfo
  {pages} {174202} (\bibinfo {year} {2014})}\BibitemShut {NoStop}%
\bibitem [{\citenamefont {Luitz}\ \emph {et~al.}(2015)\citenamefont {Luitz},
  \citenamefont {Laflorencie},\ and\ \citenamefont {Alet}}]{luitz2015}%
  \BibitemOpen
  \bibfield  {author} {\bibinfo {author} {\bibfnamefont {D.~J.}\ \bibnamefont
  {Luitz}}, \bibinfo {author} {\bibfnamefont {N.}~\bibnamefont {Laflorencie}},
  \ and\ \bibinfo {author} {\bibfnamefont {F.}~\bibnamefont {Alet}},\ }\href
  {\doibase 10.1103/PhysRevB.91.081103} {\bibfield  {journal} {\bibinfo
  {journal} {Phys. Rev. B}\ }\textbf {\bibinfo {volume} {91}},\ \bibinfo
  {pages} {081103} (\bibinfo {year} {2015})}\BibitemShut {NoStop}%
\bibitem [{\citenamefont {Vasseur}\ \emph {et~al.}(2015)\citenamefont
  {Vasseur}, \citenamefont {Parameswaran},\ and\ \citenamefont
  {Moore}}]{quantum_revivals_Vasseur_2015}%
  \BibitemOpen
  \bibfield  {author} {\bibinfo {author} {\bibfnamefont {R.}~\bibnamefont
  {Vasseur}}, \bibinfo {author} {\bibfnamefont {S.~A.}\ \bibnamefont
  {Parameswaran}}, \ and\ \bibinfo {author} {\bibfnamefont {J.~E.}\
  \bibnamefont {Moore}},\ }\href {\doibase 10.1103/PhysRevB.91.140202}
  {\bibfield  {journal} {\bibinfo  {journal} {Phys. Rev. B}\ }\textbf {\bibinfo
  {volume} {91}},\ \bibinfo {pages} {140202} (\bibinfo {year}
  {2015})}\BibitemShut {NoStop}%
\bibitem [{\citenamefont {Nandkishore}\ and\ \citenamefont
  {Huse}(2015)}]{nandkishore15review}%
  \BibitemOpen
  \bibfield  {author} {\bibinfo {author} {\bibfnamefont {R.}~\bibnamefont
  {Nandkishore}}\ and\ \bibinfo {author} {\bibfnamefont {D.~A.}\ \bibnamefont
  {Huse}},\ }\href {\doibase 10.1146/annurev-conmatphys-031214-014726}
  {\bibfield  {journal} {\bibinfo  {journal} {Annual Review of Condensed Matter
  Physics}\ }\textbf {\bibinfo {volume} {6}},\ \bibinfo {pages} {15} (\bibinfo
  {year} {2015})}\BibitemShut {NoStop}%
\bibitem [{\citenamefont {Vosk}\ \emph {et~al.}(2015)\citenamefont {Vosk},
  \citenamefont {Huse},\ and\ \citenamefont {Altman}}]{vosk15prx}%
  \BibitemOpen
  \bibfield  {author} {\bibinfo {author} {\bibfnamefont {R.}~\bibnamefont
  {Vosk}}, \bibinfo {author} {\bibfnamefont {D.~A.}\ \bibnamefont {Huse}}, \
  and\ \bibinfo {author} {\bibfnamefont {E.}~\bibnamefont {Altman}},\ }\href
  {\doibase 10.1103/PhysRevX.5.031032} {\bibfield  {journal} {\bibinfo
  {journal} {Phys. Rev. X}\ }\textbf {\bibinfo {volume} {5}},\ \bibinfo {pages}
  {031032} (\bibinfo {year} {2015})}\BibitemShut {NoStop}%
\bibitem [{\citenamefont {Agarwal}\ \emph {et~al.}(2015)\citenamefont
  {Agarwal}, \citenamefont {Gopalakrishnan}, \citenamefont {Knap},
  \citenamefont {M\"uller},\ and\ \citenamefont {Demler}}]{agarwal15grif}%
  \BibitemOpen
  \bibfield  {author} {\bibinfo {author} {\bibfnamefont {K.}~\bibnamefont
  {Agarwal}}, \bibinfo {author} {\bibfnamefont {S.}~\bibnamefont
  {Gopalakrishnan}}, \bibinfo {author} {\bibfnamefont {M.}~\bibnamefont
  {Knap}}, \bibinfo {author} {\bibfnamefont {M.}~\bibnamefont {M\"uller}}, \
  and\ \bibinfo {author} {\bibfnamefont {E.}~\bibnamefont {Demler}},\ }\href
  {\doibase 10.1103/PhysRevLett.114.160401} {\bibfield  {journal} {\bibinfo
  {journal} {Phys. Rev. Lett.}\ }\textbf {\bibinfo {volume} {114}},\ \bibinfo
  {pages} {160401} (\bibinfo {year} {2015})}\BibitemShut {NoStop}%
\bibitem [{\citenamefont {Devakul}\ and\ \citenamefont
  {Singh}(2015)}]{Singh_numerics_2015}%
  \BibitemOpen
  \bibfield  {author} {\bibinfo {author} {\bibfnamefont {T.}~\bibnamefont
  {Devakul}}\ and\ \bibinfo {author} {\bibfnamefont {R.~R.~P.}\ \bibnamefont
  {Singh}},\ }\href {\doibase 10.1103/PhysRevLett.115.187201} {\bibfield
  {journal} {\bibinfo  {journal} {Phys. Rev. Lett.}\ }\textbf {\bibinfo
  {volume} {115}},\ \bibinfo {pages} {187201} (\bibinfo {year}
  {2015})}\BibitemShut {NoStop}%
\bibitem [{\citenamefont {Schreiber}\ \emph {et~al.}(2015)\citenamefont
  {Schreiber}, \citenamefont {Hodgman}, \citenamefont {Bordia}, \citenamefont
  {L{\"u}schen}, \citenamefont {Fischer}, \citenamefont {Vosk}, \citenamefont
  {Altman}, \citenamefont {Schneider},\ and\ \citenamefont
  {Bloch}}]{schreiber15obs}%
  \BibitemOpen
  \bibfield  {author} {\bibinfo {author} {\bibfnamefont {M.}~\bibnamefont
  {Schreiber}}, \bibinfo {author} {\bibfnamefont {S.~S.}\ \bibnamefont
  {Hodgman}}, \bibinfo {author} {\bibfnamefont {P.}~\bibnamefont {Bordia}},
  \bibinfo {author} {\bibfnamefont {H.~P.}\ \bibnamefont {L{\"u}schen}},
  \bibinfo {author} {\bibfnamefont {M.~H.}\ \bibnamefont {Fischer}}, \bibinfo
  {author} {\bibfnamefont {R.}~\bibnamefont {Vosk}}, \bibinfo {author}
  {\bibfnamefont {E.}~\bibnamefont {Altman}}, \bibinfo {author} {\bibfnamefont
  {U.}~\bibnamefont {Schneider}}, \ and\ \bibinfo {author} {\bibfnamefont
  {I.}~\bibnamefont {Bloch}},\ }\href {\doibase 10.1126/science.aaa7432}
  {\bibfield  {journal} {\bibinfo  {journal} {Science}\ }\textbf {\bibinfo
  {volume} {349}},\ \bibinfo {pages} {842} (\bibinfo {year}
  {2015})}\BibitemShut {NoStop}%
\bibitem [{\citenamefont {Imbrie}(2016)}]{imbrie16proof}%
  \BibitemOpen
  \bibfield  {author} {\bibinfo {author} {\bibfnamefont {J.~Z.}\ \bibnamefont
  {Imbrie}},\ }\href {\doibase 10.1103/PhysRevLett.117.027201} {\bibfield
  {journal} {\bibinfo  {journal} {Phys. Rev. Lett.}\ }\textbf {\bibinfo
  {volume} {117}},\ \bibinfo {pages} {027201} (\bibinfo {year}
  {2016})}\BibitemShut {NoStop}%
\bibitem [{\citenamefont {Vasseur}\ and\ \citenamefont
  {Moore}(2016)}]{Vasseur_2016}%
  \BibitemOpen
  \bibfield  {author} {\bibinfo {author} {\bibfnamefont {R.}~\bibnamefont
  {Vasseur}}\ and\ \bibinfo {author} {\bibfnamefont {J.~E.}\ \bibnamefont
  {Moore}},\ }\href {http://stacks.iop.org/1742-5468/2016/i=6/a=064010}
  {\bibfield  {journal} {\bibinfo  {journal} {Journal of Statistical Mechanics:
  Theory and Experiment}\ }\textbf {\bibinfo {volume} {2016}},\ \bibinfo
  {pages} {064010} (\bibinfo {year} {2016})}\BibitemShut {NoStop}%
\bibitem [{\citenamefont {Kondov}\ \emph {et~al.}(2015)\citenamefont {Kondov},
  \citenamefont {McGehee}, \citenamefont {Xu},\ and\ \citenamefont
  {DeMarco}}]{kondov2015exp}%
  \BibitemOpen
  \bibfield  {author} {\bibinfo {author} {\bibfnamefont {S.~S.}\ \bibnamefont
  {Kondov}}, \bibinfo {author} {\bibfnamefont {W.~R.}\ \bibnamefont {McGehee}},
  \bibinfo {author} {\bibfnamefont {W.}~\bibnamefont {Xu}}, \ and\ \bibinfo
  {author} {\bibfnamefont {B.}~\bibnamefont {DeMarco}},\ }\href {\doibase
  10.1103/PhysRevLett.114.083002} {\bibfield  {journal} {\bibinfo  {journal}
  {Phys. Rev. Lett.}\ }\textbf {\bibinfo {volume} {114}},\ \bibinfo {pages}
  {083002} (\bibinfo {year} {2015})}\BibitemShut {NoStop}%
\bibitem [{\citenamefont {Choi}\ \emph {et~al.}(2016)\citenamefont {Choi},
  \citenamefont {Hild}, \citenamefont {Zeiher}, \citenamefont {Schau{\ss}},
  \citenamefont {Rubio-Abadal}, \citenamefont {Yefsah}, \citenamefont
  {Khemani}, \citenamefont {Huse}, \citenamefont {Bloch},\ and\ \citenamefont
  {Gross}}]{choi2016two}%
  \BibitemOpen
  \bibfield  {author} {\bibinfo {author} {\bibfnamefont {J.-y.}\ \bibnamefont
  {Choi}}, \bibinfo {author} {\bibfnamefont {S.}~\bibnamefont {Hild}}, \bibinfo
  {author} {\bibfnamefont {J.}~\bibnamefont {Zeiher}}, \bibinfo {author}
  {\bibfnamefont {P.}~\bibnamefont {Schau{\ss}}}, \bibinfo {author}
  {\bibfnamefont {A.}~\bibnamefont {Rubio-Abadal}}, \bibinfo {author}
  {\bibfnamefont {T.}~\bibnamefont {Yefsah}}, \bibinfo {author} {\bibfnamefont
  {V.}~\bibnamefont {Khemani}}, \bibinfo {author} {\bibfnamefont {D.~A.}\
  \bibnamefont {Huse}}, \bibinfo {author} {\bibfnamefont {I.}~\bibnamefont
  {Bloch}}, \ and\ \bibinfo {author} {\bibfnamefont {C.}~\bibnamefont
  {Gross}},\ }\href {\doibase 10.1126/science.aaf8834} {\bibfield  {journal}
  {\bibinfo  {journal} {Science}\ }\textbf {\bibinfo {volume} {352}},\ \bibinfo
  {pages} {1547} (\bibinfo {year} {2016})}\BibitemShut {NoStop}%
\bibitem [{\citenamefont {Smith}\ \emph {et~al.}(2016)\citenamefont {Smith},
  \citenamefont {Lee}, \citenamefont {Richerme}, \citenamefont {Neyenhuis},
  \citenamefont {Hess}, \citenamefont {Hauke}, \citenamefont {Heyl},
  \citenamefont {Huse},\ and\ \citenamefont {Monroe}}]{smith2016mbl}%
  \BibitemOpen
  \bibfield  {author} {\bibinfo {author} {\bibfnamefont {J.}~\bibnamefont
  {Smith}}, \bibinfo {author} {\bibfnamefont {A.}~\bibnamefont {Lee}}, \bibinfo
  {author} {\bibfnamefont {P.}~\bibnamefont {Richerme}}, \bibinfo {author}
  {\bibfnamefont {B.}~\bibnamefont {Neyenhuis}}, \bibinfo {author}
  {\bibfnamefont {P.~W.}\ \bibnamefont {Hess}}, \bibinfo {author}
  {\bibfnamefont {P.}~\bibnamefont {Hauke}}, \bibinfo {author} {\bibfnamefont
  {M.}~\bibnamefont {Heyl}}, \bibinfo {author} {\bibfnamefont {D.~A.}\
  \bibnamefont {Huse}}, \ and\ \bibinfo {author} {\bibfnamefont
  {C.}~\bibnamefont {Monroe}},\ }\href {http://dx.doi.org/10.1038/nphys3783}
  {\bibfield  {journal} {\bibinfo  {journal} {Nat Phys}\ }\textbf {\bibinfo
  {volume} {12}},\ \bibinfo {pages} {907} (\bibinfo {year} {2016})}\BibitemShut
  {NoStop}%
\bibitem [{\citenamefont {Wolf}\ \emph {et~al.}(2008)\citenamefont {Wolf},
  \citenamefont {Verstraete}, \citenamefont {Hastings},\ and\ \citenamefont
  {Cirac}}]{wolf2008area}%
  \BibitemOpen
  \bibfield  {author} {\bibinfo {author} {\bibfnamefont {M.~M.}\ \bibnamefont
  {Wolf}}, \bibinfo {author} {\bibfnamefont {F.}~\bibnamefont {Verstraete}},
  \bibinfo {author} {\bibfnamefont {M.~B.}\ \bibnamefont {Hastings}}, \ and\
  \bibinfo {author} {\bibfnamefont {J.~I.}\ \bibnamefont {Cirac}},\ }\href
  {\doibase 10.1103/PhysRevLett.100.070502} {\bibfield  {journal} {\bibinfo
  {journal} {Phys. Rev. Lett.}\ }\textbf {\bibinfo {volume} {100}},\ \bibinfo
  {pages} {070502} (\bibinfo {year} {2008})}\BibitemShut {NoStop}%
\bibitem [{\citenamefont {Verstraete}\ \emph {et~al.}(2004)\citenamefont
  {Verstraete}, \citenamefont {Garc{\'{\i}}a-Ripoll},\ and\ \citenamefont
  {Cirac}}]{verstraete04mpdo}%
  \BibitemOpen
  \bibfield  {author} {\bibinfo {author} {\bibfnamefont {F.}~\bibnamefont
  {Verstraete}}, \bibinfo {author} {\bibfnamefont {J.~J.}\ \bibnamefont
  {Garc{\'{\i}}a-Ripoll}}, \ and\ \bibinfo {author} {\bibfnamefont {J.~I.}\
  \bibnamefont {Cirac}},\ }\href@noop {} {\bibfield  {journal} {\bibinfo
  {journal} {Phys. Rev. Lett.}\ }\textbf {\bibinfo {volume} {93}},\ \bibinfo
  {eid} {207204} (\bibinfo {year} {2004})},\ \Eprint
  {http://arxiv.org/abs/cond-mat/0406426} {cond-mat/0406426} \BibitemShut
  {NoStop}%
\bibitem [{\citenamefont {Zwolak}\ and\ \citenamefont
  {Vidal}(2004)}]{zwolak04mpo}%
  \BibitemOpen
  \bibfield  {author} {\bibinfo {author} {\bibfnamefont {M.}~\bibnamefont
  {Zwolak}}\ and\ \bibinfo {author} {\bibfnamefont {G.}~\bibnamefont {Vidal}},\
  }\href {\doibase 10.1103/PhysRevLett.93.207205} {\bibfield  {journal}
  {\bibinfo  {journal} {Phys. Rev. Lett.}\ }\textbf {\bibinfo {volume} {93}},\
  \bibinfo {pages} {207205} (\bibinfo {year} {2004})}\BibitemShut {NoStop}%
\bibitem [{\citenamefont {Pirvu}\ \emph {et~al.}(2010)\citenamefont {Pirvu},
  \citenamefont {Murg}, \citenamefont {Cirac},\ and\ \citenamefont
  {Verstraete}}]{pirvu10mpo}%
  \BibitemOpen
  \bibfield  {author} {\bibinfo {author} {\bibfnamefont {B.}~\bibnamefont
  {Pirvu}}, \bibinfo {author} {\bibfnamefont {V.}~\bibnamefont {Murg}},
  \bibinfo {author} {\bibfnamefont {J.~I.}\ \bibnamefont {Cirac}}, \ and\
  \bibinfo {author} {\bibfnamefont {F.}~\bibnamefont {Verstraete}},\
  }\href@noop {} {\bibfield  {journal} {\bibinfo  {journal} {New Journal of
  Physics}\ }\textbf {\bibinfo {volume} {12}},\ \bibinfo {pages} {025012}
  (\bibinfo {year} {2010})},\ \Eprint {http://arxiv.org/abs/0804.3976}
  {0804.3976} \BibitemShut {NoStop}%
\bibitem [{\citenamefont {Pollmann}\ \emph {et~al.}(2016)\citenamefont
  {Pollmann}, \citenamefont {Khemani}, \citenamefont {Cirac},\ and\
  \citenamefont {Sondhi}}]{Pollmann_2016_MPOMBL}%
  \BibitemOpen
  \bibfield  {author} {\bibinfo {author} {\bibfnamefont {F.}~\bibnamefont
  {Pollmann}}, \bibinfo {author} {\bibfnamefont {V.}~\bibnamefont {Khemani}},
  \bibinfo {author} {\bibfnamefont {J.~I.}\ \bibnamefont {Cirac}}, \ and\
  \bibinfo {author} {\bibfnamefont {S.~L.}\ \bibnamefont {Sondhi}},\ }\href
  {\doibase 10.1103/PhysRevB.94.041116} {\bibfield  {journal} {\bibinfo
  {journal} {Phys. Rev. B}\ }\textbf {\bibinfo {volume} {94}},\ \bibinfo
  {pages} {041116} (\bibinfo {year} {2016})}\BibitemShut {NoStop}%
\bibitem [{Note1()}]{Note1}%
  \BibitemOpen
  \bibinfo {note} {One can consider generalized spin echo protocols using
  orthogonal array techniques \cite {}, but a simple application of such a
  strategy leads to an exponential number of pulses (owing to multi-body
  interactions in the phenomenological MBL Hamiltonian \cite {}).}\BibitemShut
  {Stop}%
\bibitem [{\citenamefont {Jordan}\ and\ \citenamefont
  {Wigner}(1928)}]{jordanwigner1928}%
  \BibitemOpen
  \bibfield  {author} {\bibinfo {author} {\bibfnamefont {P.}~\bibnamefont
  {Jordan}}\ and\ \bibinfo {author} {\bibfnamefont {E.}~\bibnamefont
  {Wigner}},\ }\href {\doibase 10.1007/BF01331938} {\bibfield  {journal}
  {\bibinfo  {journal} {Zeitschrift f{\"u}r Physik}\ }\textbf {\bibinfo
  {volume} {47}},\ \bibinfo {pages} {631} (\bibinfo {year} {1928})}\BibitemShut
  {NoStop}%
\bibitem [{\citenamefont {Chandran}\ \emph {et~al.}(2016)\citenamefont
  {Chandran}, \citenamefont {Pal}, \citenamefont {Laumann},\ and\ \citenamefont
  {Scardicchio}}]{chandran16dims}%
  \BibitemOpen
  \bibfield  {author} {\bibinfo {author} {\bibfnamefont {A.}~\bibnamefont
  {Chandran}}, \bibinfo {author} {\bibfnamefont {A.}~\bibnamefont {Pal}},
  \bibinfo {author} {\bibfnamefont {C.~R.}\ \bibnamefont {Laumann}}, \ and\
  \bibinfo {author} {\bibfnamefont {A.}~\bibnamefont {Scardicchio}},\ }\href
  {\doibase 10.1103/PhysRevB.94.144203} {\bibfield  {journal} {\bibinfo
  {journal} {Phys. Rev. B}\ }\textbf {\bibinfo {volume} {94}},\ \bibinfo
  {pages} {144203} (\bibinfo {year} {2016})}\BibitemShut {NoStop}%
\bibitem [{\citenamefont {Ros}\ \emph {et~al.}(2015)\citenamefont {Ros},
  \citenamefont {M{\"u}ller},\ and\ \citenamefont
  {Scardicchio}}]{ros15integrals}%
  \BibitemOpen
  \bibfield  {author} {\bibinfo {author} {\bibfnamefont {V.}~\bibnamefont
  {Ros}}, \bibinfo {author} {\bibfnamefont {M.}~\bibnamefont {M{\"u}ller}}, \
  and\ \bibinfo {author} {\bibfnamefont {A.}~\bibnamefont {Scardicchio}},\
  }\href {\doibase http://dx.doi.org/10.1016/j.nuclphysb.2014.12.014}
  {\bibfield  {journal} {\bibinfo  {journal} {Nuclear Physics B}\ }\textbf
  {\bibinfo {volume} {891}},\ \bibinfo {pages} {420 } (\bibinfo {year}
  {2015})}\BibitemShut {NoStop}%
\bibitem [{\citenamefont {Verstraete}\ \emph {et~al.}(2008)\citenamefont
  {Verstraete}, \citenamefont {Murg},\ and\ \citenamefont
  {Cirac}}]{verstraete08algo}%
  \BibitemOpen
  \bibfield  {author} {\bibinfo {author} {\bibfnamefont {F.}~\bibnamefont
  {Verstraete}}, \bibinfo {author} {\bibfnamefont {V.}~\bibnamefont {Murg}}, \
  and\ \bibinfo {author} {\bibfnamefont {J.}~\bibnamefont {Cirac}},\
  }\href@noop {} {\bibfield  {journal} {\bibinfo  {journal} {Advances in
  Physics}\ }\textbf {\bibinfo {volume} {57}},\ \bibinfo {pages} {143}
  (\bibinfo {year} {2008})},\ \Eprint {http://arxiv.org/abs/0907.2796}
  {0907.2796} \BibitemShut {NoStop}%
\bibitem [{\citenamefont {Mazza}\ \emph {et~al.}(2013)\citenamefont {Mazza},
  \citenamefont {Rizzi}, \citenamefont {Lukin},\ and\ \citenamefont
  {Cirac}}]{mazza12majorana}%
  \BibitemOpen
  \bibfield  {author} {\bibinfo {author} {\bibfnamefont {L.}~\bibnamefont
  {Mazza}}, \bibinfo {author} {\bibfnamefont {M.}~\bibnamefont {Rizzi}},
  \bibinfo {author} {\bibfnamefont {M.~D.}\ \bibnamefont {Lukin}}, \ and\
  \bibinfo {author} {\bibfnamefont {J.~I.}\ \bibnamefont {Cirac}},\ }\href
  {\doibase 10.1103/PhysRevB.88.205142} {\bibfield  {journal} {\bibinfo
  {journal} {Phys. Rev. B}\ }\textbf {\bibinfo {volume} {88}},\ \bibinfo
  {pages} {205142} (\bibinfo {year} {2013})}\BibitemShut {NoStop}%
\bibitem [{Note2()}]{Note2}%
  \BibitemOpen
  \bibinfo {note} {Strictly speaking, owing to a symmetry of our Hamiltonian,
  total magnetization $\DOTSB \sum@ \slimits@ _i S_i^z$ is a conserved
  quantity. Hence, the leading order correction to $\sigma _z^{[0]}$ in $l$-bit
  basis need not be $\tau _x^{[0]}$. Here, we focus on a generic situation
  independent of symmetry considerations for the purpose of understanding
  qualitative behaviors.}\BibitemShut {Stop}%
\bibitem [{\citenamefont {Nielsen}\ and\ \citenamefont
  {Chuang}(2011)}]{nielsenchuang}%
  \BibitemOpen
  \bibfield  {author} {\bibinfo {author} {\bibfnamefont {M.~A.}\ \bibnamefont
  {Nielsen}}\ and\ \bibinfo {author} {\bibfnamefont {I.~L.}\ \bibnamefont
  {Chuang}},\ }\href@noop {} {\emph {\bibinfo {title} {Quantum Computation and
  Quantum Information: 10th Anniversary Edition}}},\ \bibinfo {edition} {10th}\
  ed.\ (\bibinfo  {publisher} {Cambridge University Press},\ \bibinfo {address}
  {New York, NY, USA},\ \bibinfo {year} {2011})\BibitemShut {NoStop}%
\bibitem [{\citenamefont {Daley}\ \emph {et~al.}(2012)\citenamefont {Daley},
  \citenamefont {Pichler}, \citenamefont {Schachenmayer},\ and\ \citenamefont
  {Zoller}}]{Daley:2012bd}%
  \BibitemOpen
  \bibfield  {author} {\bibinfo {author} {\bibfnamefont {A.~J.}\ \bibnamefont
  {Daley}}, \bibinfo {author} {\bibfnamefont {H.}~\bibnamefont {Pichler}},
  \bibinfo {author} {\bibfnamefont {J.}~\bibnamefont {Schachenmayer}}, \ and\
  \bibinfo {author} {\bibfnamefont {P.}~\bibnamefont {Zoller}},\ }\href@noop {}
  {\bibfield  {journal} {\bibinfo  {journal} {Physical Review Letters}\
  }\textbf {\bibinfo {volume} {109}},\ \bibinfo {pages} {020505} (\bibinfo
  {year} {2012})}\BibitemShut {NoStop}%
\bibitem [{\citenamefont {Pichler}\ \emph {et~al.}(2013)\citenamefont
  {Pichler}, \citenamefont {Bonnes}, \citenamefont {Daley}, \citenamefont
  {L{\"a}uchli},\ and\ \citenamefont {Zoller}}]{Pichler:2013bs}%
  \BibitemOpen
  \bibfield  {author} {\bibinfo {author} {\bibfnamefont {H.}~\bibnamefont
  {Pichler}}, \bibinfo {author} {\bibfnamefont {L.}~\bibnamefont {Bonnes}},
  \bibinfo {author} {\bibfnamefont {A.~J.}\ \bibnamefont {Daley}}, \bibinfo
  {author} {\bibfnamefont {A.~M.}\ \bibnamefont {L{\"a}uchli}}, \ and\ \bibinfo
  {author} {\bibfnamefont {P.}~\bibnamefont {Zoller}},\ }\href@noop {}
  {\bibfield  {journal} {\bibinfo  {journal} {New Journal of Physics}\ }\textbf
  {\bibinfo {volume} {15}},\ \bibinfo {pages} {063003} (\bibinfo {year}
  {2013})}\BibitemShut {NoStop}%
\bibitem [{\citenamefont {{Islam, Rajibul}}\ \emph {et~al.}(2015)\citenamefont
  {{Islam, Rajibul}}, \citenamefont {{Ma, Ruichao}}, \citenamefont {{Preiss,
  Philipp M}}, \citenamefont {{Tai, M Eric}}, \citenamefont {{Lukin,
  Alexander}}, \citenamefont {{Rispoli, Matthew}},\ and\ \citenamefont
  {{Greiner, Markus}}}]{Islam:2015cm}%
  \BibitemOpen
  \bibfield  {author} {\bibinfo {author} {\bibnamefont {{Islam, Rajibul}}},
  \bibinfo {author} {\bibnamefont {{Ma, Ruichao}}}, \bibinfo {author}
  {\bibnamefont {{Preiss, Philipp M}}}, \bibinfo {author} {\bibnamefont {{Tai,
  M Eric}}}, \bibinfo {author} {\bibnamefont {{Lukin, Alexander}}}, \bibinfo
  {author} {\bibnamefont {{Rispoli, Matthew}}}, \ and\ \bibinfo {author}
  {\bibnamefont {{Greiner, Markus}}},\ }\href@noop {} {\bibfield  {journal}
  {\bibinfo  {journal} {Nature}\ }\textbf {\bibinfo {volume} {528}},\ \bibinfo
  {pages} {77} (\bibinfo {year} {2015})}\BibitemShut {NoStop}%
\bibitem [{Note3()}]{Note3}%
  \BibitemOpen
  \bibinfo {note} {Notice that the maximally mixed states are also the thermal
  states at infinite temperature, so that this should be the final value of the
  mutual information if the system would thermalize in the case of $X\pm $
  initial states. In our system, total polarization $\DOTSB \sum@ \slimits@ _i
  S_z^{[i]}$ is conserved. Thus, the maximum entropy of the initial state $Z\pm
  $ has to be modified accordingly. These corrections vanish in thermodynamical
  limit. See Appendix~\ref {app:XY}}\BibitemShut {NoStop}%
\bibitem [{\citenamefont {McCulloch}(2007)}]{mcculloch2007}%
  \BibitemOpen
  \bibfield  {author} {\bibinfo {author} {\bibfnamefont {I.~P.}\ \bibnamefont
  {McCulloch}},\ }\href {http://stacks.iop.org/1742-5468/2007/i=10/a=P10014}
  {\bibfield  {journal} {\bibinfo  {journal} {Journal of Statistical Mechanics:
  Theory and Experiment}\ }\textbf {\bibinfo {volume} {2007}},\ \bibinfo
  {pages} {P10014} (\bibinfo {year} {2007})}\BibitemShut {NoStop}%
\end{thebibliography}%

\end{document}